\documentclass[12pt]{article}
\usepackage{amsmath}
\usepackage{url}
\usepackage{graphicx,psfrag,epsf}
\usepackage{enumerate}
\usepackage{natbib}

\usepackage{multirow}
\usepackage{epstopdf,gensymb}
\usepackage{graphics}
\usepackage{graphicx,color}
\usepackage{amsfonts, animate,subfigure}
\usepackage{amssymb}
\usepackage[margin=1in]{geometry}
\RequirePackage[colorlinks,citecolor=blue,urlcolor=blue]{hyperref}
\usepackage{todonotes}
\usepackage{setspace}
\newcommand{\erdos}{Erd\H{o}s-R\'enyi }
\usepackage{indentfirst}
\usepackage{float}
\usepackage{amsfonts}
\usepackage{bm,pbox}
\usepackage{multirow} 
\usepackage{color}
\usepackage[margin=1in]{geometry}
\usepackage{framed}
\usepackage{amssymb,amsthm,amsfonts,amsbsy,latexsym,dsfont}
\usepackage{enumerate}
\usepackage[ansinew]{inputenc}
\usepackage{lscape}
\usepackage{rotating}
\usepackage{hyperref}
\usepackage{lmodern} 

\usepackage{tikz}
\usetikzlibrary{arrows}
 
\begin{document}
	\def\spacingset#1{\renewcommand{\baselinestretch}%
	{#1}\small\normalsize} \spacingset{1}
	\title{\bf Monitoring communication outbreaks among an unknown team of actors in dynamic networks\hspace{.2cm}}
	  \author{Ross Sparks$^1$ and James D. Wilson$^2$}
		
	  \maketitle
\begin{abstract}
This paper investigates the detection of communication outbreaks among a small team of actors in time-varying networks. We propose monitoring plans for known and unknown teams based on generalizations of the exponentially weighted moving average (EWMA)  statistic. For unknown teams, we propose an efficient neighborhood-based search to estimate a collection of candidate teams. This procedure dramatically reduces the computational complexity of an exhaustive search. Our procedure consists of two steps: communication counts between actors are first smoothed using a multivariate EWMA strategy. Densely connected teams are identified as candidates using a neighborhood search approach. These candidate teams are then monitored using a surveillance plan derived from a generalized EWMA statistic. Monitoring plans are established for collaborative teams, teams with a dominant leader, as well as for global outbreaks. We consider weighted heterogeneous dynamic networks, where the expected communication count between each pair of actors is potentially different across pairs and time, as well as homogeneous networks, where the expected communication count is constant across time and actors. Our monitoring plans are evaluated on a test bed of simulated networks as well as on the U.S. Senate co-voting network, which models the Senate voting patterns from 1857 to 2015. Our analysis suggests that our surveillance strategies can efficiently detect relevant and significant changes in dynamic networks. 
\end{abstract}

\vfill
	\footnotetext[1]{Digital Productivity,
CSIRO. Private Bag 17. North Ryde,
Sydney NSW 1670, Australia \\ \url{Ross.Sparks@csiro.au}}
	\footnotetext[2]{Department of Mathematics and Statistics,
	University of San Francisco. 
	San Francisco, CA 94117-1080 \url{jdwilson4@usfca.edu}}
\newpage
\spacingset{1.45} 

\section{Introduction}
In many applications, it is of interest to identify anomalous behavior among the actors in a time-varying network. For example in online social networks, sudden increased communications often signify illegal behavior such as fraud or collusion \citep{pandit2007netprobe, savage2014anomaly}. Anomalous changes like these are reflected by local structural changes in the network. The goal of network monitoring is to provide a surveillance plan that can detect such structural changes. Network monitoring techniques have been successfully utilized in a number of applications, including the identification of central players in terrorist groups \citep{krebs2002mapping, reid2005collecting, porter2012self}, and the detection of fraud in online networks \citep{chau2006detecting, pandit2007netprobe, akoglu2013anomaly}. As available data has become more complex, there has been a recent surge of interest in the development and application of scalable network monitoring methodologies (see \citet{savage2014anomaly} and \citet{woodall2016overview} for recent reviews).

In this paper, we investigate monitoring the interactions of a fixed collection of $n$ actors $[n] = \{1, \ldots, n\}$ over discrete times $t = 1, \ldots, T$. In general, an interaction is broadly defined and may represent, for example, communications in an online network \citep{prusiewicz2008multi}, citations in a co-authorship network \citep{liu2005co}, or gene-gene interactions in a biological network \citep{parker2015network}. We model the interactions of these actors at time $t$ by a $n \times n$ stochastic adjacency matrix $Y_t = (y_{i,j,t})$, where $y_{i,j,t}$ is the discrete random variable that represents the number of interactions between actor $i$ and actor $j$ at time $t$. Our goal is to develop a surveillance strategy to detect communication outbreaks among a subset of actors $\Omega_t \subseteq [n]$ at time $t$. 

The identification of outbreaks among a subset of actors $\Omega_t$ corresponds to detecting sudden increases in the collection of edges $\{y_{i,j,t}: i,j \in \Omega_t\}$. When the team is unknown, monitoring can be computationally expensive due to the need for identifying candidate teams. For example, consider a simple case where we know the size of the target team is $ n_{\Omega_t} = |\Omega_t|$. An exhaustive monitoring of all teams of size $n_{\Omega_t}$ requires a procedure of complexity ${n \choose n_{\Omega_t}} \approx n^{n_{\Omega_t}}$, which is infeasible even for moderately sized networks. As social networks are generally large, e.g. $n$ is on the order of 1 million for online networks like those represesenting Facebook or Twitter, exhaustive searches are not practical in real-time. To address this challenge, we propose a computationally efficient local surveillance strategy that monitors the interactions of densely connected neighborhoods through time. Our proposed strategy has computational complexity of order $n^2$, and provides a viable strategy for large networks.

Our surveillance procedure consists of two steps, which can be briefly described as follows. First, we smooth the communication counts across all pairs and time using a multivariate adaptation of the exponentially weighted moving average (EWMA) technique for smoothing Poisson counts. By monitoring the smoothed counts, our strategy is robust to sudden random oscillations in the observed count process. Next, candidate teams are identified locally for each node using a neighborhood-based approach. In particular, at time $t$ we define a candidate team for node $i \in [n]$ as one that contains larger than expected communication. Surveillance plans for these candidate teams are developed using appropriate generalizations of the multivariate EWMA  statistic.

We develop surveillance plans using the above technique in general for {\it heterogeneous} dynamic networks $\bm{Y} = \{Y_1, \ldots, Y_T\}$, where we suppose that the expected communication counts are possibly different for each pair and time, namely, $\mathbb{E}[y_{i,j,t}] = \lambda_{i,j,t}$. We consider three situations describing the team $\Omega_t$:

\begin{itemize}
	\item[(i)] \emph{Collaborative teams}: members of $\Omega_t$ communicate with one another far more than they communicate with actors outside of the team.
	\item[(ii)] \emph{Dominant leader teams}: the members of $\Omega_t$ have a dominant leader $\nu$ who communicates frequently with members of $\Omega_t$, but the members of $\Omega_t$ themselves do not necessarily communicate frequently amongst themselves. 
	\item[(iii)] \emph{Global outbreaks}: the entire network undergoes a communication outbreak, namely $\Omega_t \equiv [n]$.
\end{itemize}

Scenarios (i) and (ii) are considered for both unknown and known teams. Each of the scenarios are also considered for homogeneous networks, where $\mathbb{E}[y_{i,j,t}] \equiv \lambda$. By investigating both a test bed of simulated networks as well as a real network describing the U.S. Senate voting patterns, we find that our surveillance strategy can efficiently and reliably detect significant changes in dynamic networks.

\subsection{Related Work}

The most closely related work to our current manuscript is that introduced in \citet{heard2010bayesian}. In that paper, the authors also consider monitoring changes in communication volume between subgroups of targeted people over time. Their approach evaluates pairwise communication counts and determines whether these have significantly increased using a p-value, which assesses the deviation of the communication rate at time $t$ and what is considered normal behavior. Here, normal behavior is modeled using conjugate Bayesian models for the discrete-valued time series of communications up to time $t$. While their focus is detecting changes on the entire network, our approach considers detecting communication outbreaks for members of a small team within the dynamic network. 

There are other model-based network monitoring approaches that have been recently developed, which we briefly describe here. \citet{azarnoush2016monitoring} proposed a longitudinal logistic model that describes the (binary) occurence of an edge at time $t$ as a function of time-varying edge attributes in the sequence of networks $\bm{G}([n], T)$. Likelihood ratio tests of the fitted model are used to identify significant changes in $\bm{G}([n], T)$. \citet{peel2014detecting} developed a generalized hierarchical random graph model (GHRG) to model $\bm{G}([n], T)$. To detect anomalies, the authors used the GHRG as a null model to compare observed graphs in $\bm{G}([n], T)$ via a Bayes factor, which is calculated using bootstrap simulation. \citet{wilson2016modeling} proposed modeling and estimating change in a sequence of networks using the dynamic degree-corrected stochastic block model (DCSBM). In that work, maximum likelihood estimates of the DCSBM are used for monitoring via Shewhart control charts. Our model is similar to the DCSBM in that edges are modeled as having discrete-valued edge-weights, which flexibly model communications in social networks.

The EWMA control chart is a popular univariate monitoring technique. The multivariate EWMA process that we use here is a generalization of the univariate EWMA strategies for Poisson counts considered in \citet{weiss2007controlling, weiss2009ewma}, \citet{sparks2009improving, sparks2010early}, and \citet{zhou2012likelihood}. A related multivariate EWMA control chart has previously been successfully applied to space-time monitoring of crime \citep{zeng2004comparative, kim2008bootstrap, neill2009expectation, nakaya2010visualising}. 

Our specified dynamic network model for $\bm{Y} = \{Y_1, \ldots, Y_T\}$ is related to several well-studied random graph models, which are ubiquitous in social network analysis. For example, when $y_{i,j,t}$ are independent and identically distributed $\text{Poisson}(\lambda)$ random variables, the graph at time $t$ is an \erdos random graph model with edge connection probability $\lambda$ \citep{erdos1960evolution}. On the other hand, when $y_{i,j,t}$ are independent $\text{Poisson}(\lambda_{i,j,t})$ random variables, graph $t$ is a weighted variant of the Chung-Lu random graph model \citep{aiello2000random}. Random graph models play an important role in the statistical analysis of relational data. \citet{goldenberg2010survey} provides a recent survey about random graph models and their applications.

\subsection{Organization of this Paper}
The remainder of this paper is organized as follows. In Section \ref{sec:EWMA}, we describe how to smooth the observed communication counts using multivariate EWMA smoothing. In Section \ref{sec:surveillance} we
develop surveillance strategies for communication outbreaks among small teams of actors in a dynamic network when the target team is known. We consider collaborative teams, dominant leader teams, as well as global outbreaks. Section \ref{sec:estimating} describes our proposed local search and monitoring approach for unknown target teams. Section \ref{sec:simulations} investigates the performance of our surveillance strategies on a test-bed of simulated networks. We make recommendations on designing the plans in such a way to minimize false discovery. In Section \ref{sec:application}, we further assess the performance of our strategy by applying the plans to the heterogeneous network describing the U.S. Senate voting patterns from the 35th to the 113th Congress. We conclude with a summary of our findings and discuss directions for future work in Section \ref{sec:discussion}.

\section{Temporal EWMA Smoothing of Interactions}\label{sec:EWMA}
Throughout this work, we are concerned with detecting significant increases in communication among the members of some subset of actors $\Omega_t \subseteq [n]$. Such fluctations correspond to sudden spikes in the collection of edge weights $\{y_{i,j,t}: i,j \in \Omega_t\}$. In many cases, the communication counts $\{y_{i,j,t}: i,j \in [n], t = 1, \ldots, T\}$ are prone to random fluctuations that arise from noise in the observed process. If not accounted for, direct monitoring of counts may lead to false discovery. To reduce this possibility, we smooth the observed counts using a reflective EWMA strategy \citep{gan1993exponentially}. 

To begin, we first obtain a collection of smoothed values $\{\widetilde{y}_{i,j,t}: i,j \in [n], t = 1, \ldots, T\}$ using an EWMA strategy. Fix $\alpha \in [0,1]$, and define

\begin{equation}\label{eq:EWMA1}\widetilde{y}_{i,j,t} = \alpha ~ y_{i,j,t} + (1-\alpha)~ \widetilde{y}_{i,j,t-1}. \end{equation}

\noindent Denote the expected value of $\widetilde{y}_{i,j,t}$ by $\widetilde{\lambda}_{i,j,t}$. The expected values of these smoothed counts can be calculated using the following recursion

\begin{equation*}\label{eq:EWMA2} \widetilde{\lambda}_{i,j,t} = \alpha ~ \lambda_{i,j,t} + (1-\alpha) ~ \widetilde{\lambda}_{i,j,t-1}.\end{equation*}

\noindent In the above recursion, the initial values are set as $\widetilde{y}_{i,j,0} = \widetilde{\lambda}_{i,j,0} = {\lambda}_{i,j,1}$. Here, $\alpha$ acts as a smoothing parameter that dictates the temporal memory retained in the stochastic process $\{\widetilde{y}_{i,j,t}: i,j \in [n], t = 1, \ldots, T\}$. Large values of $\alpha$ retain less memory and result in less smoothing. In our applications, we fix $\alpha$ to 0.075 based on the previous analysis and suggestion of \citet{sparks2014detection}.

Notably, the EWMA in (\ref{eq:EWMA1}) will not reflect a change in the observed count process in the scenario that $y_{i,j,t}$ decreases immediately before a significant (anomalous) increase. To avoid this worst-case scenario, we use the reflective boundary EWMA process $\{y_{i,j,t}^*: i, j \in [n], t = 1, \ldots, T\}$, defined by

\begin{equation}\label{eq:ref_EWMA} {y}^*_{i,j,t} = \max(\alpha~\widetilde{y}_{i,j,t} + (1-\alpha)~y_{i,j,t}^*, \widetilde{\lambda}_{i,j,t}) \end{equation}

The reflective boundary EWMA specified in (\ref{eq:ref_EWMA}) is robust to sudden oscillations in the count process. Our surveillance plans will utilize the smoothed counts from (\ref{eq:ref_EWMA}) rather than the originally observed counts.

\section{Monitoring a Known Team of Actors}\label{sec:surveillance}
We begin by considering the simplest case when the target team $\Omega_t$ is known {\it a priori}. This scenario arises, for example, in the surveillance of the communication among a known active group of terrorists in a large terrorist network. We develop surveillance plans for collaborative and dominant leader teams, as well as global changes, where the entire network undergoes a communication outbreak. For each of these scenarios we describe monitoring a homogeneous sequence of networks $\bm{Y}$, where the collection of expected communications $\{\lambda_{i,j,t}: i,j \in [n], t = 1,\ldots, T\}$ are such that $\lambda_{i,j,t} \equiv \lambda$ for all $i,j$ and $t$, and further describe how to extend the plans in this regime to the more general heterogeneous case, where expected communications are possibly different accross time and actor pairs. 

In both this section and Section \ref{sec:estimating}, we will make use of two tunable parameters -- $\alpha \in [0,1]$: a smoothing parameter that controls the extent to which a proposed EWMA statistic has temporal memory, and $h(\cdot, \cdot)$: threshold functions that are chosen to control false discovery of the proposed monitoring plan. We fix $\alpha = 0.075$ based on previous analysis conducted in \citet{sparks2014detection}. The threshold functions $h(\cdot, \cdot)$ are chosen via simulation of the monitored process. We describe how these are chosen in detail in the Appendix. 

Throughout this and the following section, let $\widetilde{y}_{i,j,t}$ and $y^*_{i,j,t}$ be the EWMA and reflective boundary EWMA defined in (\ref{eq:EWMA1}) and (\ref{eq:ref_EWMA}), respectively. Further, we denote $n_{\Omega_t} = |\Omega_t|$ as the number of individuals in the team.
\subsection{$\Omega_t$ is a Collaborative Team}\label{sec:GEWMA}
We first consider monitoring for outbreaks among a collaborative team $\Omega_t$, wherein all members of $\Omega_t$ are expected to communicate regularly. An outbreak in a collaborative team is reflected by a large average number of communications between members $i, j \in \Omega_t$. To detect such outbreaks, we analyze the mean, $\mu_{\Omega_t}$, of the smoothed interactions in the collection defined as
 
\begin{equation}\label{eq:mean}\mu_{\Omega_t} = \mathbb{E}\left[\sum_{i \in \Omega_t} \sum_{j \in \Omega_t} \widetilde{y}_{i,j,t}\right] = \sum_{i \in \Omega_t} \sum_{j \in \Omega_t} \widetilde{\lambda}_{i,j,t}\end{equation} 

In the case that $\bm{Y}$ is homogeneous, note that $\mu_{\Omega_t} = n_{\Omega_t}^2 \lambda$. We use a group - EWMA (GEWMA) statistic to identify outbreaks among the actors in $\Omega_t$. The $\text{GEWMA}_t$ process is defined by the following recursion

\begin{equation}\label{eq:GEWMA} \text{GEWMA}_t = \max\left(\alpha \sum_{i \in \Omega_t} \sum_{j \in \Omega_t} \widetilde{y}_{i,j,t}+ (1-\alpha) ~ \text{GEWMA}_{t-1}, ~ \mu_{\Omega_t}\right), \end{equation}
	
\noindent where the initial value $\text{GEWMA}_1 = \sum_{i \in \Omega_t} \sum_{j \in \Omega_t} \widetilde{y}_{i,j,1}$. 


For homogeneous networks, we use the $\text{GEWMA}_t$ process from (\ref{eq:GEWMA}) and flag an outbreak within the team $\Omega_t$ when

\begin{equation}\label{eq:GEWMA_flag3} \sqrt{\text{GEWMA}_t} - n_{\Omega_t}\sqrt{\lambda} > h_G(\lambda, n_{\Omega_t}),\end{equation}

\noindent where $h_{G}(n_{\Omega_t}, \lambda)$ is designed to give the plan a low false discovery rate. Importantly, the square root transform of the $\text{GEWMA}_t$ process in (\ref{eq:GEWMA_flag3}) stabilizes the variance of the process to a constant value (see \citet{bartlett1936square}). Thus, the left hand side of (\ref{eq:GEWMA_flag3}) is no longer a function of the mean $\lambda$. Indeed, we find from simulation that the threshold $h_{G}(n_{\Omega_t}, \lambda)$ is \emph{not} a function of $\lambda$; hence, even in the heterogeneous case we can use a plan with the threshold $h_G(n_{\Omega_t})$. We describe how to choose the value $h_G(n_{\Omega_t})$ in the Appendix. Thus for heterogeneous networks, we flag an outbreak in the team $\Omega_t$ when
	
\begin{equation}\label{eq:GEWMA_flag2} \sqrt{\text{GEWMA}_t} - \sqrt{\sum_{i \in \Omega_t} \sum_{j \in \Omega_t} \widetilde{\lambda}_{i,j,t}} > h_G(n_{\Omega_t}). \end{equation}

In practice, a target team $\Omega_t$ may purposefully reduce their communication levels prior to, say, planning a crime, which may hamper early detection when using the $\text{GEWMA}_t$ statistic defined in (\ref{eq:GEWMA}). To avoid this scenario, one can alternatively use a reflective boundary GEWMA statistic defined as

\begin{equation}\label{eq:GEWMA_ref} \text{GEWMA}^*_t = \sum_{i \in \Omega_t}\sum_{j \in \Omega_t} {y}^*_{i,j,t},\end{equation}

\noindent and apply an analogous plan as defined in (\ref{eq:GEWMA_flag2}).

\subsection{$\Omega_t$ Has a Dominant Leader}\label{sec:DEWMA}
We now consider the scenario in which the target team $\Omega_{t}$ has a known dominant leader $\nu \in [n]$. We expect that $\nu$ will have a high level of communication with the members of $\Omega_{t}$, but unlike the collaborative team setting, the members of $\Omega_{t}$ do not neccessarily significantly interact with one another. In this case, an outbreak is signalled when there is either a significant rate of communications between $\nu$ and the members of $\Omega_{t}$, or by a significant rate of interactions among the members of $\Omega_t$. As we primarily need to be concerned with the communications between a single actor and a collection of actors, we develop a monitoring strategy that exploits sparsity in the interactions among the members of $\Omega_{t}$. At time $t$, we monitor only the collection of actors that (a) significantly communicate with the dominant leader $\nu$, and (b) significantly communicate with one another. That is, we identify the dominant leader team $\Omega_{t}$ by following two steps. First we identify the team $W_{\nu,t}$ that contains all individuals in $[n]$ with a significant number of interactions with $\nu$, namely

\begin{equation}\label{eq:DEWMA_known_team1}W_{\nu, t} = \{i \neq \nu \in [n]: \sqrt{y^{*}_{\nu,i,t} + y^*_{i, \nu, t}} - \sqrt{\widetilde{\lambda}_{\nu,i,t} + \widetilde{\lambda}_{i,\nu,t}} > k\}.\end{equation}

Next we refine the team $W_{\nu, t}$ to include only those members who share a significant number of communications. We set

\begin{equation}\label{eq:DEWMA_known_team2}{\Omega}_{t} = \{i,j \in W_{\nu, t}: \sqrt{{y}^*_{i,j,t}} - \sqrt{\widetilde{\lambda}_{i,j,t}} > k ~\text{or}~ \sqrt{{y}^*_{j,i,t}} - \sqrt{\widetilde{\lambda}_{j,i,t}} > k \}.\end{equation}
	
The value $k$ is a suitable constant that helps identify members of the target group and is chosen to control the size of the team $\Omega_{t}$. We consider the choice of $k$ in our simulation study in Section \ref{sec:simulations}. To monitor $\Omega_{t}$, we use the dominant leader EWMA (DEWMA) statistic, defined as


\begin{equation}\label{eq:DEWMA} \text{DEWMA}_{\nu,t} = \sum_{i \in W_{\nu, t}}  \left(y^*_{i,\nu,t} + y^*_{\nu, i, t}\right) + \sum_{i \in \Omega_t}\sum_{j \in \Omega_t} y^*_{i, j, t}. \end{equation}
	
When $\nu$ is known, we can use the DEWMA statistic from (\ref{eq:DEWMA}) to flag outbreaks in a dominant leader team. In the case that $\bm{Y}$ is homogeneous, we flag an outbreak when

\begin{equation}\label{eq:DEWMA_flag2} \sqrt{\text{DEWMA}_{\nu, t}} - \sqrt{2n_{W_{\nu, t}}\lambda + n_{\Omega_t}^2\lambda} > h_D(n_{ \lambda, \Omega_{t}}, \lambda). \end{equation}
	
Above, $h_D(n_{\Omega_t}, \lambda)$ is chosen to control false discovery. Once again simulations suggest that the square root transformation rids the dependence of the threshold $h_D(n_{\Omega_t}, \lambda)$ on $\lambda$. Thus, we use the following general surveillance plan for heterogeneous networks when $\nu$ is known
	
\begin{equation}\label{eq:DEWMA_flag} \sqrt{\text{DEWMA}_{\nu,t}} - \sqrt{\sum_{i \in W_{\nu, t}}  \left(\widetilde{\lambda}_{i,\nu,t} + \widetilde{\lambda}_{\nu, i, t}\right) + \sum_{i \in \Omega_t}\sum_{j \in \Omega_t} \widetilde{\lambda}_{i, j, t}} > h_D(n_{\Omega_t}), \end{equation}

We note that when the team and dominant leader are both unknown, the plan in (\ref{eq:DEWMA_flag}) is complicated by the fact that we must estimate $\nu$ and $\Omega_t$. We discuss our strategy to handle this in Section \ref{sec:estimating}.

\subsection{Global Outbreaks}
We now consider the case when there is a significant increase in the number of interactions among every pair of actors in the network, i.e., when $\Omega_t \equiv [n]$ for all $t$. One can generally detect this anomaly early by monitoring the aggregated interactions over the target network. To monitor the network for a global outbreak, one can directly extend the $\text{GEWMA}_t$ statistic from (\ref{eq:GEWMA}) to the entire network. Note that in the case that $\Omega_t \equiv [n]$, we have from (\ref{eq:mean}) that $\mu_{[n]} =\sum_{i \in [n]} \sum_{j \in [n]} \widetilde{\lambda}_{i,j,t}$. Following our previous development of the $\text{GEWMA}_t$ statistic in (\ref{eq:GEWMA}), we define the total-EWMA (TEWMA) statistic using the following recursion

\begin{equation}\label{eq:TEWMA} \text{TEWMA}_t = \max\left(\alpha \sum_{i \in [n]} \sum_{j \in [n]} \widetilde{y}_{i,j,t}+ (1-\alpha) ~ \text{TEWMA}_{t-1}, ~ \mu_{[n]}\right), \end{equation}

\noindent where $\text{TEWMA}_1 = \sum_{i \in [n]} \sum_{j \in [n]} \widetilde{y}_{i,j,1}$, and $\alpha \in [0,1]$ is chosen to smooth the TEWMA process. Using the statistic in (\ref{eq:TEWMA}), we flag a {global outbreak} in homogeneous networks when 

\begin{equation}\label{eq:TEWMA_flag2} \sqrt{\text{TEWMA}_t} - n\sqrt{\lambda} > h_T(\lambda, n).\end{equation}
	
The threshold $h_{T}(n, \lambda)$ designed to give the plan a low enough false discovery rate, and is chosen in the same manner as plan (\ref{eq:GEWMA_flag3}).	As before, $h_T(n, \lambda)$ does not depend on the expected communication counts due to the square root transform. Hence, in general we flag an outbreak in heterogeneous networks when 
	
\begin{equation}\label{eq:TEWMA_flag3} \sqrt{\text{TEWMA}_t} - \sqrt{\sum_{i \in [n]} \sum_{j \in [n]} \widetilde{\lambda}_{i,j,t}} > h_T(n).\end{equation}

To avoid issues arising from sudden oscillations in counts, we can instead use the reflected-boundary TEWMA statistic
 
\begin{equation} \text{TEWMA}^*_t = \sum_{i \in [n]} \sum_{j \in [n]} y^*_{i,j,t}, \end{equation} 
\noindent and apply the plan given in (\ref{eq:TEWMA_flag3}).

\section{Monitoring of an Unknown Team of Actors} \label{sec:estimating}
In many applications, $\Omega_t$ is \emph{not} known {\it a priori}. In this situation, there are two primary difficulties that one must address. First, the unknown team must be efficiently estimated. An exhaustive search for an anomalous team has complexity of order $n^{n_{\Omega_t}}$; thus, it is important to employ scalable approaches for estimation. When $\Omega_t$ is known, the $\text{GEWMA}_t$ and $\text{DEWMA}_{\nu,t}$ statistics are invariant to variations in the communication means. However, when $\Omega_t$ is unknown these statistics are no longer invariant to heterogeneous communication rates through time. Thus the second complication comes in adapting the monitoring plan for a changing mean in heterogeneous networks. In this section we describe a local search strategy to identify densely connected teams on which our proposed statistics can be used for monitoring. Since the global outbreak plan in (\ref{eq:TEWMA_flag3}) is invariant to mean changes, we only need to consider the scenarios when $\Omega_t$ is either a collaborative team or a dominant leader team.

\subsection{Estimating Unknown Teams}
Here, we describe our local search strategy to estimate collaborative teams as well as teams with a dominant leader. 
\subsubsection{Collaborative Teams}
When the target team is unknown and collaborative, we propose monitoring a collection of densely connected teams $\bm{\Omega}_{C,t} := \{\widehat{\Omega}_{\ell, t}: \ell \in [n]\}$ at each time $t$. We define a candidate team $\widehat{\Omega}_{\ell, t}$ as one in which all constituent members significantly interact. In particular, for each $\ell \in [n]$ and each time $t$, we identify the candidate team

\begin{equation}\label{eq:collab_estimate} \widehat{\Omega}_{\ell, t} = \{i \in [n]: \sqrt{y^*_{i,\ell, t}} - \sqrt{\widetilde{\lambda}_{i,\ell,t}} > k, \text{or} ~ \sqrt{y^*_{\ell,i, t}} - \sqrt{\widetilde{\lambda}_{\ell, i, t}} > k\}.\end{equation}

\noindent Above, $k$ is a suitable constant with good detection properties and is chosen via simulation. Our specification of each candidate team $\widehat{\Omega}_{\ell, t}$ is motivated by empirical properties of real networks. One can view $\widehat{\Omega}_{\ell, t}$ structurally as a hub with center node $\ell$. Hub structures commonly arise in sparse social and biological networks as well as the well-studied scale-free family of networks \citep{barabasi1999emergence, tan2014learning}. Thus if the unknown team is suspected to be a collaborative team, we propose monitoring at most $n$ densely connected teams.

\subsubsection{Dominant Leader Teams}
When the dominant leader $\nu$ and target team $\Omega_t$ is unknown, we monitor a collection of candidate dominant leader teams  $\bm{\Omega}_{D,t} := \{\widehat{\Omega}_{\nu, t}: \nu \in [n]\}$ at each time $t$. Like the identification of dominant leader teams in Section \ref{sec:surveillance}, we identify a collection of candidate dominant leader teams that have a significantly large rate of communication. First for a fixed leader $\nu \in [n]$ we identify a team $\widehat{W}_{\nu,t}$ by finding all individuals in $[n]$ with a significant number of interactions with $\nu$ given by

\begin{equation}\label{eq:DEWMA_team1}\widehat{W}_{\nu, t} = \{i \neq \nu \in [n]: \sqrt{y^{*}_{\nu,i,t} + y^*_{i, \nu, t}} - \sqrt{\widetilde{\lambda}_{\nu,i,t} + \widetilde{\lambda}_{i,\nu,t}} > k\}\end{equation}

We next refine the team $\widehat{W}_{\nu, t}$ to include only those members who share a significant number of interactions. Namely, we specify the team $\widehat{\Omega}_{\nu, t}$ as

\begin{equation}\label{eq:DEWMA_team2}\widehat{\Omega}_{\nu, t} = \{i,j \in \widehat{W}_{\nu, t}: \sqrt{{y}^*_{i,j,t}} - \sqrt{\widetilde{\lambda}_{i,j,t}} > k ~\text{or}~ \sqrt{{y}^*_{j,i,t}} - \sqrt{\widetilde{\lambda}_{j,i,t}} > k \}\end{equation}
	
The value $k$ is a suitable constant that helps identify members of the target group with larger than expected communications with the dominant leader $\nu$. We note that rather than a normal standardized score to identify $\Omega_{t}$, we use a \lq signal-to-noise\rq \, team identification scheme in (\ref{eq:DEWMA_team1}) as this strategy can efficiently avoid unusual changes that involve very low communication levels. 


\subsection{Adapting the Plans for Heterogeneous Networks}
Once the candidate teams $\bm{\Omega}_{C, t} = \{\widehat{\Omega}_{\ell, t}: \ell \in [n]\}$ and $\bm{\Omega}_{D, t} = \{\widehat{\Omega}_{\nu, t}: \nu \in [n]\}$ have been estimated for each time $t$, we can develop a monitoring plan. For $\ell, \nu \in [n]$, define the following local GEWMA and DEWMA statistics

\begin{equation}\label{eq:GEWMA_local}
	\text{GEWMA}^*_{\ell, t} = \sum_{i \in \widehat{\Omega}_{\ell, t}}\sum_{j \in \widehat{\Omega}_{\ell, t}} y^*_{i,j,t}
\end{equation}

\begin{equation}\label{eq:DEWMA_local}
	\text{DEWMA}^*_{\nu, t} = \sum_{i \in \widehat{W}_{\nu, t}}  \left(y^*_{i,\nu,t} + y^*_{\nu, i, t}\right) + \sum_{i \in \widehat{\Omega}_{\nu, t}}\sum_{j \in \widehat{\Omega}_{\nu, t}} y^*_{i, j, t}.
\end{equation}

When the observed network is homogeneous, one can readily monitor collaborative and dominant leader teams by using plans (\ref{eq:GEWMA_flag2}) and (\ref{eq:DEWMA_flag2}), respectively, for the local GEWMA and DEWMA statistics in (\ref{eq:GEWMA_local}) and (\ref{eq:DEWMA_local}). When the network is heterogeneous, we develop an adaptive plan for surveillance as follows. Note that for a fixed candidate collaborative team $\widehat{\Omega}_{\ell, t}$, the plan in (\ref{eq:GEWMA_flag2}) can be re-expressed as

\begin{equation}\label{eq:flag_GEWMA}
	\sqrt{\text{GEWMA}^*_{\ell, t} / h_G^2(\lambda, n_{\widehat{\Omega}_{\ell, t}})} - \sqrt{\sum_{i \in \widehat{\Omega}_{\ell, t}}\sum_{j \in \widehat{\Omega}_{\ell, t}}\lambda_{i,j,t} / h_G^2(\lambda, n_{\widehat{\Omega}_{\ell, t}})} > 1
\end{equation}

\noindent Importantly the threshold in plan (\ref{eq:flag_GEWMA}) no longer depends on the observed data. We exploit this property and define an adaptive plan using the local adaptive group-EWMA (AGEWMA) statistic:

\begin{equation}\text{AGEWMA}_{\ell, t} =  \text{GEWMA}^*_{\ell, t} / h_G^2(\widetilde{\lambda}_{i,j,t}, n_{\widehat{\Omega}_{\ell, t}}).\end{equation}

\noindent For an unknown team $\Omega_t$, a communication outbreak is flagged when

\begin{equation}\label{eq:AGEWMA}
	\sqrt{\text{AGEWMA}_{\ell, t}} - \sqrt{\sum_{i \in \widehat{\Omega}_{\ell, t}}\sum_{j \in \widehat{\Omega}_{\ell, t}}\widetilde{\lambda}_{i,j,t} / h_G^2(\widetilde{\lambda}_{i,j,t}, n_{\widehat{\Omega}_{\ell, t}})} > 1,
\end{equation}

\noindent for any $\ell \in [n]$. Here, the team must be re-estimated at each time period $t$. This adaptive plan in (\ref{eq:AGEWMA}) has the same in-control ATS value used to design the homogeneous plans for all $\lambda_{i,j,t}$.

We can use a similar adaptive plan to identify communication outbreaks in candidate dominant leader teams. Define the local adaptive dominant leader - EWMA (ADEWMA) statistic by

\begin{equation}
	\text{ADEWMA}_{\nu, t} = \sum_{i \in \widehat{W}_{\nu, t}}\left(\frac{y_{i,\nu,t}^*}{h_D(\widetilde{\lambda}_{i,\nu, t}, n_{\widehat{\Omega}_{\nu, t}})}  + \frac{y_{j,\nu,t}^*}{h_D(\widetilde{\lambda}_{j,\nu, t}, n_{\widehat{\Omega}_{\nu, t}})}\right) + \sum_{i \in \widehat{\Omega}_{\nu, t}}\sum_{j \in \widehat{\Omega}_{\nu, t}}\frac{y_{i,j,t}^*}{h_D(\widetilde{\lambda}_{i, j, t}, n_{\widehat{\Omega}_{\nu, t}})}.
\end{equation}

Using an analagous argument as above for the adaptive GEWMA plan, we flag a communication outbreak among dominant leader teams when

\begin{align}\label{eq:ADEWMA}
	\sqrt{\text{ADEWMA}_{\nu, t}} &- \sqrt{\sum_{i \in \widehat{W}_{\nu, t}}\left(\frac{\widetilde{\lambda}_{i,\nu,t}^*}{h_D(\widetilde{\lambda}_{i,\nu, t}, n_{\widehat{\Omega}_{\nu, t}})}  + \frac{\widetilde{\lambda}_{j,\nu,t}^*}{h_D(\widetilde{\lambda}_{j,\nu, t}, n_{\widehat{\Omega}_{\nu, t}})}\right) + \sum_{i \in \widehat{\Omega}_{\nu, t}}\sum_{j \in \widehat{\Omega}_{\nu, t}}\frac{\widetilde{\lambda}_{i,j,t}^*}{h_D(\widetilde{\lambda}_{i, j, t}, n_{\widehat{\Omega}_{\nu, t}})}}\nonumber  \\ 
	> 1 \end{align}

\noindent for any $\nu \in [n]$. There are two distinct scenarios in which an outbreak will be flagged by the plan (\ref{eq:ADEWMA}). In the first scenario, an outbreak is detected if the team size of any candidate team significantly increases. This is likely to happen when, for instance, a leader of an organized crime is trying to recruit a team. In the second scenario, an outbreak is detected when the number of interactions within any candidate team significantly increases. This can occur in two ways: (i) when individuals within the same team interact more with individuals outside of their current group, or (ii) members of the group interact significantly more frequently among themselves. Combinations of (i) and (ii) may also flag communication outbreaks.

\section{Simulation Study}\label{sec:simulations}
We now access the utility of our proposed surveillance plans on a test bed of simulated networks. We consider two types of communication outbreaks among small target teams. In the first scenario, we simulate a collaborative team outbreak, where every actor in a small and unknown team is involved in the outbreak. In the second scenario, the target team has an unknown dominant leader whose communication levels with the remaining team undergoes an outbreak. For each of these cases, we investigate the effectiveness of the GEWMA and DEWMA strategies.

For each simulation, we generate 100 in-control networks followed by 500 networks that have undergone an outbreak. We record the time to signal - the number of networks after the change until a signal is flagged - of the DEWMA and GEWMA plans and repeat the experiment 10000 times for the collaborative team outbreak and 1000 times for the dominant leader outbreak. To evaluate the performance of a plan, we record the average time to signal (ATS) over the collection of simulations. We present the results for all simulations in Tables 1 - 12 in the Appendix.

\subsection{Collaborative Team Outbreaks}\label{sec:collaborative}

Tables 1 through 10 outline the detection properties of simulated collaborative team outbreaks for networks of size $n = 100$. To simulate an outbreak, we select a fixed but hidden team $\Omega \subseteq \{1, \ldots, 100\}$. In the first 100 in-control networks, communication counts among the nodes in $\Omega$ have mean $\lambda$. In the remaining networks, the nodes in $\Omega$ have an increased mean communication count of $(1+\delta)\lambda$. We simulate networks with target teams of size $n_{\Omega} = 6, 7, 8, 9,$ and 10. For each time series of networks, we estimate candidate collaborative teams and dominant leader teams via (\ref{eq:collab_estimate}) and (\ref{eq:DEWMA_team2}) and then apply the GEWMA and DEWMA plans from (\ref{eq:GEWMA_flag3}) and (\ref{eq:DEWMA_flag2}), respectively.

\subsubsection{The $\text{GEWMA}_t$ Plan}

In the first part of our study, we simulate homogeneous target networks with mean communication counts of either $\lambda = 0.20$ or $0.70$. We investigate significance thresholds $k$ between 0.05 and 0.40 in increments of 0.05. Table 1 explores changes in communication counts in a team of size 6. Table 1 reveals that $k = 0.40$ provides the best performance for both $\lambda$ values.

We extend the first simulation to seek the best plan for detecting the collaborative team $\Omega$, when $n_\Omega = 6$ and $n = 100$. We investigate significance thresholds of $k$ between $0.40$ and $0.70$ for expected communication rates of $\lambda = 0.20, 0.40$ and $0.70$. Together, Tables 1 and 2 indicate that $k = 0.60$ is the best choice for all $\lambda$ and $\Omega$ involving 6 of the 100 actors. Furthermore we find that the performance of the GEWMA plan strongly depends on an appropriate choice of $k$; the detection performance of the GEWMA plan is dramatically improved for $k = 0.60$. 

We repeat the collaborative team outbreak simulation for a target team of size 7, 8, 9, and 10. In each simulation, we seek the best significance threshold $k$ for homogeneous networks with mean communication $\lambda = 0.20, 0.40$ and $0.70$. We report the ATS over 10000 simulations for each of these settings in Tables 3 - 6. Our results suggest that $k = 0.50$ is the best choice for all $\lambda$ when $n_\Omega$ is 8, 9, or 10, while $k = 0.50$ or 0.60 is most suitable for networks where the target team is of size 7. This result suggests that there is an inverse relationship between the optimal value of $k$ and the size of the target team. This is helpful in deciding the choice of $k$ for the GEWMA plan, and it appears that $k = 0.50$ is a robust choice for the outbreaks considered in this study. 

\subsubsection{The $\text{DEWMA}_{\nu,t}$ Plan}

Tables 7 - 10 report the results of the DEWMA surveillance plan on the collaborative team outbreaks described above for target teams of size 6, 7, 8, and 9. For each setting, $k = 0.45$ tends to be the best choice for significance threshold. The only exception is in the case that the team is of size 9 and the mean communication is $\lambda = 0.70$, in which case $k = 0.40$ is the better choice. 

\subsubsection{Comparison of the $\text{GEWMA}_t$ and $\text{DEWMA}_{\nu,t}$ Plans} 

In comparing the results for the GEWMA and DEWMA plans on the collaborative team outbreak simulation, we find that in general the GEWMA plan outperforms the DEWMA plan. In particular, the GEWMA strategy detects the collaborative team sooner than its counterpart. For example when $\delta=1$ and $\lambda=0.2$, the strategy based on $\text{GEWMA}_t$ in Table 2 had an ATS equal to 11.62 ($k = 0.60$) whereas the technology based on $\text{DEWMA}_{\nu,t}$ in Table 7 had an ATS equal to 12.90 ($k = 0.45$). Similarly, when $\delta=0.50$ and $\lambda=0.70$; the $\text{GEWMA}_t$ strategy had an ATS equal to 8.54 ($k = 0.50$); whereas, the $\text{DEWMA}_{\nu,t}$ plan had an ATS of 8.87 ($k=0.40$). 

\subsubsection{Is the Methodology fit-for-purpose?}
In order to judge whether the technology is fit for purpose we consider the monitoring of a crime. To be effective, we would like our strategy to flag the planning of a crime within seven days. We assume the following specifications of team behavior:
 
\begin{enumerate}
\item In order to plan a crime, team members should call each other at least 0.5 per day during the planning phase. We consider this to be the lowest level of communication necessary to plan a crime.
\item The planning stage of the crime would result in at least a doubling of their usual
communication intensity during this planning stage.
\item The usefulness specification is that detection should be well within 7 days of the start (i.e., the out-of-control ATS $<$ 7).
\end{enumerate}
 
The last specification allows law enforcement agencies enough time for appropriate detective work to be carried out and potentially avoid catastrophic events such as terrorism. The optimal plan for $\lambda = 0.4$ and 0.7 pass the usefulness test by flagging within seven days on average for all groups (e.g., with $\lambda = 0.4$, k = 0.6 the $\text{GEWMA}_t$ statistics detect the outbreak on average in 6.93 days). On the other hand, when the overall  communication in the network is relatively sparse ($\lambda = 0.2$), this fit for purpose test is only met for collaborative teams having 8 or more members. 

\subsection{Dominant Leader Team Outbreaks}

\begin{figure}[h]\label{fig:dominant_leader_sims}
	\centering
\begin{tabular}{c c}
	{\bf Simulation 1} & {\bf Simulation 2} \\
	& \\
\begin{tikzpicture}
\begin{scope}[every node/.style={circle,thick,draw}]
    \node (1) at (0,0) {1};
    \node (3) at (0,3) {3};
    \node (2) at (5,3) {2};
    \node (6) at (2.5,1) [circle, thick, draw, fill = black!20]{6};
    \node (5) at (5,0) {5};
    \node (4) at (2.5,4) {4} ;
\end{scope}

\begin{scope}[>={stealth},
              every edge/.style={draw=black,very thick}]
    \path [->] (2) edge node {} (5);
    \path [->] (4) edge node {} (1);
    \path [->] (6) edge node {} (1);
    \path [->] (6) edge node {} (2);
    \path [->] (6) edge node {} (3);
    \path [->] (6) edge node {} (4);
    \path [->] (6) edge node {} (5);
\end{scope}
\end{tikzpicture}
&
\begin{tikzpicture}
\begin{scope}[every node/.style={circle,thick,draw}]
    \node (1) at (0,0) {1};
    \node (3) at (0,3) {3};
    \node (2) at (5,3) {2};
    \node (6) at (2.5,1) [circle, thick, draw, fill = black!20]{6};
    \node (5) at (5,0) {5};
    \node (4) at (2.5,4) {4};
	\node (7) at (2.5, -.5) {7};
\end{scope}

\begin{scope}[>={stealth},
              every edge/.style={draw=black,very thick}]
    \path [->] (2) edge node {} (5);
    \path [->] (4) edge node {} (1);
	\path [->] (7) edge node {} (5);
	\path [->] (4) edge[bend left = 60] node {} (7);
    \path [->] (6) edge node {} (1);
    \path [->] (6) edge node {} (2);
    \path [->] (6) edge node {} (3);
    \path [->] (6) edge node {} (4);
    \path [->] (6) edge node {} (5);
	\path [->] (6) edge node {} (7);
\end{scope}
\end{tikzpicture}\\
& \\
{\bf Simulation 3} & {\bf Simulation 4} \\
& \\
\begin{tikzpicture}
\begin{scope}[every node/.style={circle,thick,draw}]
    \node (1) at (0,0) {1};
    \node (3) at (0,3) {3};
    \node (2) at (5,3) {2};
    \node (6) at (2.5,1) [circle, thick, draw, fill = black!20]{6};
    \node (5) at (5,0) {5};
    \node (4) at (2.5,4) {4};
	\node (7) at (2.5, -.5) {7};
	\node (8) at (1.26, -.75) {8};
\end{scope}

\begin{scope}[>={stealth},
              every edge/.style={draw=black,very thick}]
    \path [->] (2) edge node {} (5);
    \path [->] (4) edge node {} (1);
	\path [->] (7) edge node {} (5);
	\path [->] (4) edge[bend left = 60] node {} (7);
    \path [->] (6) edge node {} (1);
    \path [->] (6) edge node {} (2);
    \path [->] (6) edge node {} (3);
    \path [->] (6) edge node {} (4);
    \path [->] (6) edge node {} (5);
	\path [->] (6) edge node {} (7);
	\path [->] (6) edge node {} (8);
	\path [->] (7) edge node {} (8);
	\path [->] (4) edge node {} (8);
	\path [->] (1) edge node {} (8);
\end{scope}
\end{tikzpicture}
&
\begin{tikzpicture}
\begin{scope}[every node/.style={circle,thick,draw}]
    \node (1) at (0,0) {1};
    \node (3) at (0,3) {3};
    \node (2) at (5,3) {2};
    \node (6) at (2.5,1) [circle, thick, draw, fill = black!20]{6};
    \node (5) at (5,0) {5};
    \node (4) at (2.5,4) {4};
	\node (7) at (2.5, -.5) {7};
	\node (8) at (1.25, -.75) {8};
	\node (9) at (3.75, 3.5) {9};
\end{scope}

\begin{scope}[>={stealth},
              every edge/.style={draw=black,very thick}]
    \path [->] (2) edge node {} (5);
    \path [->] (4) edge node {} (1);
	\path [->] (7) edge node {} (5);
	\path [->] (4) edge[bend left = 60] node {} (7);
    \path [->] (6) edge node {} (1);
    \path [->] (6) edge node {} (2);
    \path [->] (6) edge node {} (3);
    \path [->] (6) edge node {} (4);
    \path [->] (6) edge node {} (5);
	\path [->] (6) edge node {} (7);
	\path [->] (6) edge node {} (8);
	\path [->] (7) edge node {} (8);
	\path [->] (4) edge node {} (8);
	\path [->] (1) edge node {} (8);
	\path [->] (6) edge node {} (9);
	\path [->] (3) edge node {} (9);
	\path [->] (9) edge node {} (2);
	\path [->] (8) edge[bend left = 40] node {} (9);
\end{scope}
\end{tikzpicture}
\end{tabular}
\caption{Dominant leader target teams for the simulation study. Teams are of size 6, 7, 8, and 9 among a network of size 100. For each simulation, a communication outbreak occurs only on the directed edges shown. In each simulation, node 6 is the dominant leader and communicates with every member of the team.}
\end{figure}
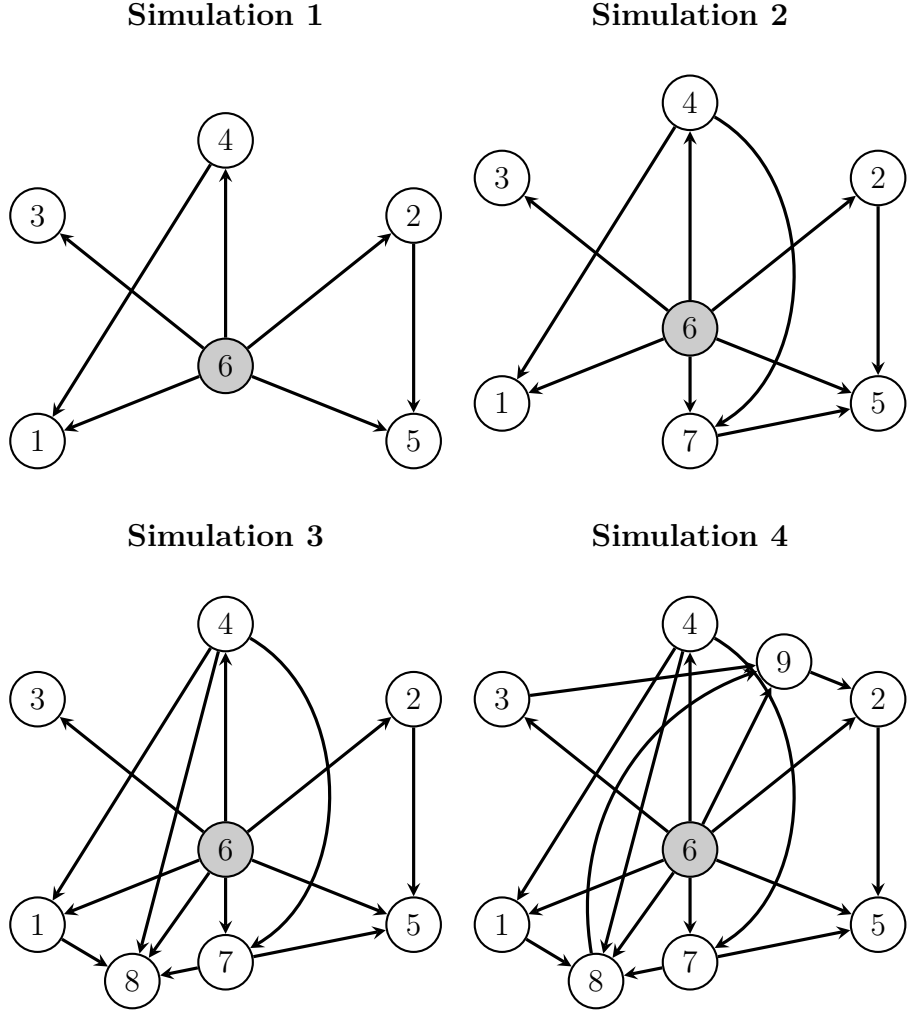

We now investigate the performance of the GEWMA and DEWMA plans when the outbreak occurs among a fixed but unknown dominant leader team in a homogeneous dynamic network. We simulate the networks with the same specifications as the collaborative team study in Section \ref{sec:collaborative}, except now the outbreak only occurs on a fixed subset of communications in the team (rather than throughout the entire team as in the collaborative team scenario). In particular, we consider four different dominant leader teams where a communication outbreak occurs on the directed edges shown in Figure \ref{fig:dominant_leader_sims}. In each of these four teams, team member 6 is assumed to be the dominant leader and communicates with all other members of the team.

We assess the performance of the GEWMA and DEWMA plans on these dominant leader outbreaks and report the results in Tables 11 and 12. Our results suggest that again the choice of $k$ plays an important role in establishing the best performing monitoring strategy. Furthermore, across all values of $\lambda$ $k$, and $n_{\Omega}$, we found that the DEWMA method outperformed the GEWMA strategy in this simulation study. Both methods witness improved performance as the signal to noise ratio ($\delta$) increases. Our results provide empirical evidence that the DEWMA plan is an effective strategy when the target team has a dominant leader, or when the team is more sparsely connected than a collaborative team.

\subsection{Heterogeneous Networks with no Outbreak}
We now assess the performance of the ADEWMA plan from (\ref{eq:ADEWMA}) on heterogeneous networks that undergo no outbreak, but whose size changes through time. Without loss of generality, we fix the mean communication count between node $i$ and $j$ at time $t$ as $\lambda_{i,j,t} = a \vert i - j \vert + 0.90,$ for a fixed constant $a < 0$. This specification gives a higher likelihood of communication between nodes that are close to one another in the ordering of the nodes. To vary the size of the network through time, we fix lower ($m_L$) and upper bounds ($m_H$) and select the size of the $t^{\text{th}}$ network $n_t$ by randomly drawing a discrete value uniformly from the interval $[m_L, m_H]$. 

As there is no outbreak in our simulated collection of networks, we seek a plan that identifies no change for some fixed number of time steps. By investigating this aspect of the ADEWMA plan, we can better understand how to control the number of false discoveries under a null model where no outbreak is present. For our current study, we seek an ADEWMA plan that delivers an ATS of 100. We note that one could alternatively seek an ATS of 370 to match the standard three sigma strategy of Shewhart control charts, but the choice is arbitrary. We vary the values of $a$, $m_L$, and $m_H$ and identify the threhold adjustment that acquires the desired ATS over 1000 simulations. The threshold adjustments and calculated ATS are provided in Table 13. 

The simulation results in Table 13 reveal that the ADEWMA plan with threshold 0.984 has an in-control ATS closest to the desired value of 100 when $m_{H} > 135$. On the other hand, when $m_{H} \le 135$ selecting a threshold of $1$ delivers the best plan. These results suggest that the ADEWMA plan is robust to large changes in the size of the network from one time to the next. In many applications (like our application in Section \ref{sec:application}), the size $n_t$ is likely to have a small variation over time. We find that in these situations the ADEWMA plan witnesses an improvement in overall robustness.

\section{Application to U.S. Congressional Voting}\label{sec:application}
We now apply the GEWMA monitoring plan from (\ref{eq:GEWMA_flag2}) to investigate the dynamic relationship between Republican and Democratic senators in the U.S. Congress. We analyze the voting habits of each U.S. senator according to his or her vote (yay, nay, or abstain) on each bill that went to Congress. We investigate these voting habits from 1857 (Congress 35) to 2015 (Congress 113). 

We generated a dynamic network to model the co-voting patterns among U.S. Senators in the following manner. We first collected the raw roll call voting data for each bill from \url{http://voteview.com}. For each Congress, we generate a new network, where the senators of that Congress are the nodes, and the edge weight between two senators is the number of bills for which those two senators voted concurrently in that Congress. We restrict our analysis to Republican and Democrat senators only (thus ignoring the Independent party and other affiliations).

Predictable behavior is regarded as in-control. To model in-control behavior, we use a logistic regression model to predict whether two senators will vote the same on a newly submitted bill. We fit a logistic model to estimate the probability that a senator (Senator A) would vote the same as another senator (Senator B) using the following predictors: (a) the political affiliation of each senator (Senators A and B), (b) which party had a majority in the Congress, (c) the proportion of that majority, and (d) the proportion of representation of Senator A's political affiliation. The expected number of votes from Senator A to Senator B was calculated by multiplying the predicted probability from the logistic regression by the total number of votes for that senator. This count was assumed to be Poisson distributed with in-control mean given by this expected count.  

In this application we are interested in both unusually high counts and unusually low counts. Therefore we run two one-sided charts. In particular, for a target team $\Omega_t$ we analyze the $\text{GEWMA}_t$ statistic from (\ref{eq:GEWMA}), as well as the lower GEWMA ($\text{L-GEWMA}_t$) statistic defined by

$$\text{L-GEWMA}_t = \min(\alpha ~ \sum_{i \in \Omega_t}\sum_{j \in \Omega_t} \widetilde{y}_{i,j,t}+ (1-\alpha) ~\text{L-GEWMA}_{t-1}, \mu_{\Omega_t}),$$

\noindent where $\alpha$ was fixed to be 0.075. The plans are trained using simulation to deliver an in-control false alarm rate of 200. The GEWMA and L-GEWMA curves were calculated from two sources (i) the likelihood of Republicans voting with Democrats, and (ii) the likelihood of Democrats voting with Republicans. We do not expect our co-voting patterns to remain in-control and predictable; thus, we are particularly interested in identifying sustained periods of unusual behavior.

\begin{figure}[ht]\label{fig:Congress_application}
	\centering
\begin{tabular}{c}
	{\bf Democrat Propensity to Vote with Republicans} \\
	\includegraphics[width = 0.75\textwidth]{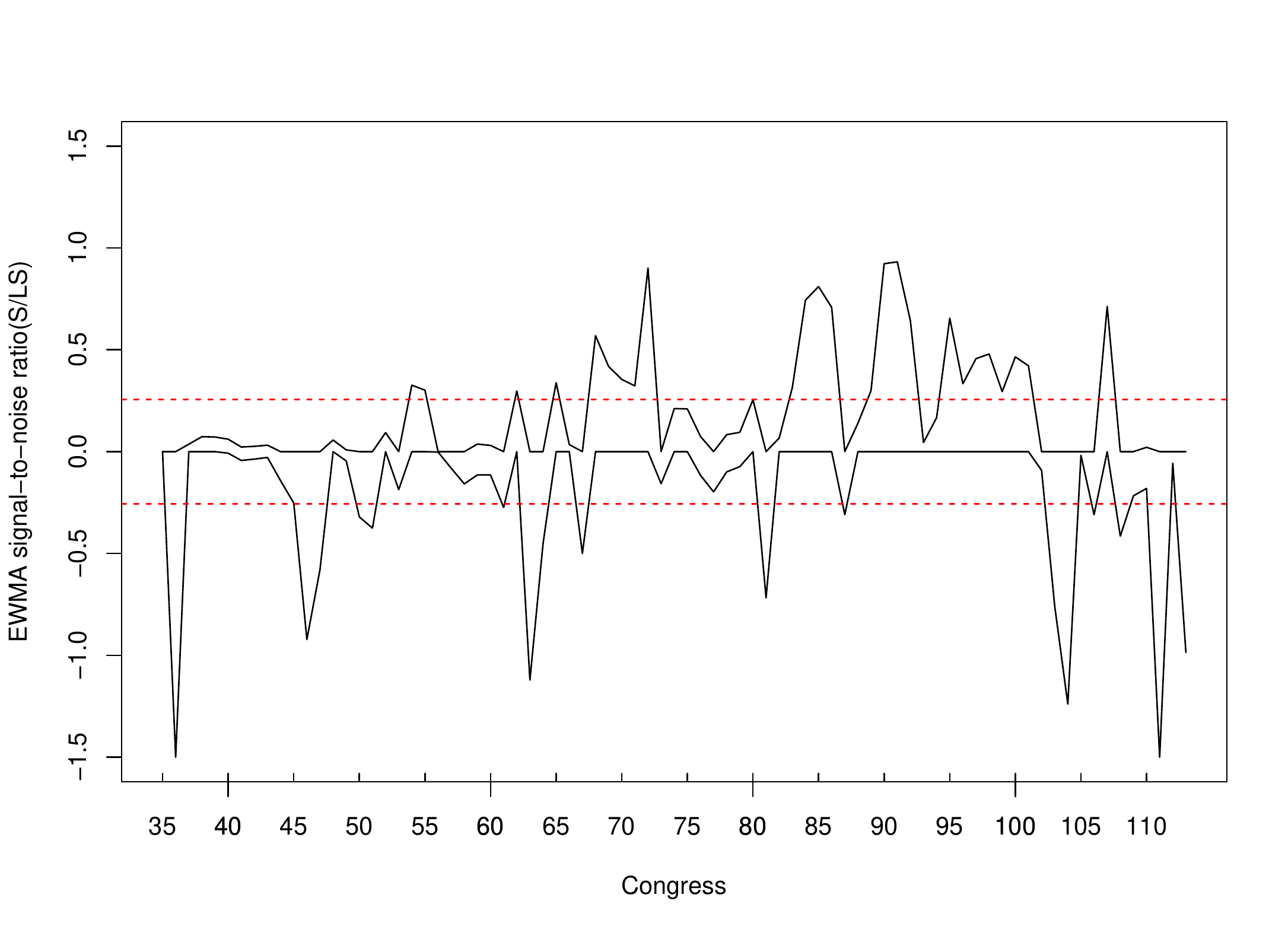}\\
	\\
	{\bf Republican Propensity to Vote with Democrats}\\
	 \includegraphics[width = 0.75\textwidth]{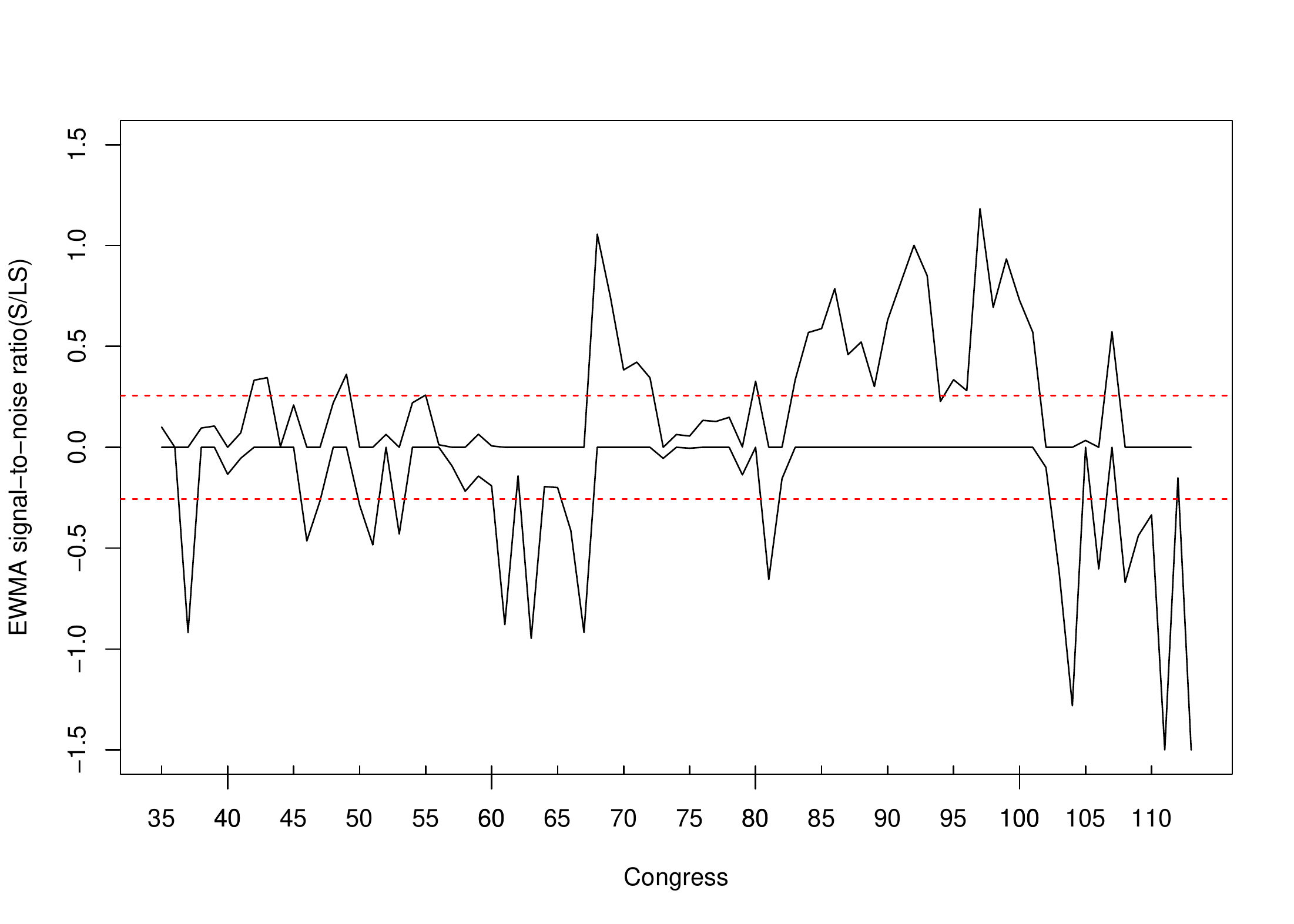}
\end{tabular}
\caption{GEWMA and L-GEWMA control charts for monitoring (TOP): the likelihood of Democratic senators to vote with Republican senators, and (BOTTOM): the likelihood of Republican senators to vote with Democratic senators. Red dotted lines mark the control limits of the GEWMA signal to noise value for each Congress. In each plot, the upper curve represents the GEWMA statistic and the lower curve represents the L-GEWMA statistic over time.}
\end{figure}

The GEWMA and L-GEWMA curves are plotted in Figure \ref{fig:Congress_application}. These plots reveal several interesting trends in the Congressional co-voting network. First, the tendency for Republican and Democratic senators to vote with one another has been significantly low beginning from Congress 103. This finding supports the political polarization theory observed in \citet{moody2013portrait}, who noted that the Republican and Democrat schism began around the time of Bill Clinton's first term as president (Congress 103). Second, there was a sustained coherence of voting between opposing political parties between Congress 85 (1957) and Congress 100 (1987). During this time, the likelihood of one party concurrently voting with the other opposing party was significantly high. Much of this time period coincides with the so-called ``Rockefeller Republican'' era (1960 - 1980) in which Republican party members were known to hold particularly moderate views like the former governor of New York, Nelson Rockefeller \citep{rae1989decline, smith2014his}. This finding was also identified using network surveillance techniques in \citet{wilson2016modeling}. 

\section{Discussion}\label{sec:discussion}

This paper introduces novel and computationally feasible surveillance plans for identifying communication outbreaks in dynamic networks. In the worst-case scenario when the target team is unknown, the proposed method monitors at most $n^2$ candidate teams, which dramatically improves the computational memory needed for an exhaustive search. Our new plan uses a general multivariate EWMA approach to accumulate temporal memory of communication counts. The approach can easily be extended to situtations with more than one communication channel. Plans were extended to handle networks with heterogeneous mean counts (as in the application) and the value of our proposed plans was further demonstrated with simulated applications.

In our simulation study, we found that our new approach is able to effectively identify 

\clearpage
\clearpage

\noindent outbreaks even when the outbreak covers a small number of communications ($< 1\%$ of total communications). These results suggest that the technology will be particularly useful in crime management as crime is typically committed by gangs of a small size \citep{a2014juvenile}. Furthermore, we believe that law enforcement agencies would value our proposed technique as it could be used to help gain insights on persons of interest, e.g., it could be applied juvenile crime rings as a preventative tool to help reduce repeat offenders.  

We found that when the outbreak is global across all communications of the targeted people, using the $\text{TEWMA}_t$ plan is the best approach and this plan is invariant of the distribution of communication counts in the target network. If the communication outbreaks involves a small sub-group of the targeted people then the group-EWMA ($\text{GEWMA}_t$) plan has best performance. As the size of the outbreak group is seldom known in advance, applying these plans simultaneously in a single plan may offer a more robust means to detect the full range of potential outbreaks. 

Our proposed technique motivates several areas of future research. For example, future work should explore the potential of extending this approach to cover geographic dimensions (see \citet{carley2013near}) to account for the spatial nature of observed dynamic systems. Furthermore, one can explore other ways of estimating the target team for monitoring. New approaches could involve defining people in the targeted network with either increased connectivity or historically a high connectivity. The target group itself could be regarded as varying according to whether they achieve a certain level of connectivity with the leaders, or average connectivity within the target group. In principle, one could also estimate teams of individuals that are most densely connected at time $t$ using a community detection or extraction algorithm on the network $Y_t$ \citep{lancichinetti2010statistical, zhao2011community, wilson2014testing}. Alternatively, one could identify candidate teams in a network with statistically significant edges using a p-value technique like that developed in \citet{wilson2013measuring}. 

Finally, this paper arbitrarily selected the temporal smoothing parameter $\alpha = 0.075$. Therefore future research effort could be devoted to selecting an appropriate value for the multivariate temporal smoothing. We believe that this effort should be devoted either to establishing an appropriate robust choice for $\alpha$, or to alternatively varying the choice of $\alpha$ for each communication count so as to exploit local trends in the network such as the work done in \citet{capizzi2003adaptive}.

\section*{Appendix}

\subsection*{Specification of Threshold Values}
Simulation methods were used to estimate the thresholds for the $\text{DEWMA}_{\nu,t}$ and $\text{GEWMA}_t$ plans so as to deliver an in-control ATS of approximately 100. The thresholds for both the collaborative team and the dominant leader team were established in the identical manner. To avoid redundancy, we will describe the simulation procedure to determine thresholds in the collaborative team scenario.

For the $\text{DEWMA}_{\nu,t}$ plan, we simulated networks of size $n = 100,125,150,\dots,375, 400$. For each network, we fixed the temporal memory as $\alpha = 0.10$ and generated homogeneous networks with mean counts equal to $\lambda = 0.01,0.02,0.03,\dots,0.10,0.15,0.20,\dots,0.95,1.0$. For each combination, the thresholds $h_D(\lambda, n)$ are estimated to obtain the fixed ATS. These values were then used to build the following regression model: 

\begin{align*}\log(h_D(\lambda, n)) &= \beta_0 + \beta_1 n +\beta_2 n^2 +\beta_3 n^3+\beta_4 \lambda +\beta_5 \lambda^2 +
\beta_6~\mathbb{I}(\lambda<0.95) + \beta_7~\mathbb{I}(\lambda<0.95)\lambda \\
& +\beta_8  \log(\lambda)+\beta_9 n \log(\lambda)
                  +\beta_{10} n \lambda+\beta_{11} n\lambda^2 + \text{error}.
									\end{align*}
									
%

Once fitted, the above regression model was used to estimate the thresholds for the $\text{DEWMA}_{\nu,t}$ plan for homogeneous networks with mean count $\lambda$ and size $n$. The above fitted model delivers an in-control ATS within $100 \pm 15$ for the range of $100 \le n \le 400$, $0.01 \le \lambda \le 1.0$ and $\alpha = 0.10$.  The standard error of the model was 0.0043 and the correlation between the model fitted values and the corresponding actual simulated $h_D(\lambda, n)$ values was 0.9996.

For the $\text{GEWMA}_t$ plan, we estimated the threshold $h_G(\lambda, n)$ in a similar way as above. We generated networks of size $n = 100, 125, 150,\dots,975, 1000$, fixed $\alpha=0.10$, and simulated homogeneous networks with mean counts $\lambda= 0.01,0.02,0.3,\dots,0.1,0.15,0.2,\dots,0.95,1.0$. For each combination, we estimated the threshold $h_G(\lambda, n)$ through simulation, and then used these estimates to build the following regression model:

\begin{align*} 1 / h_G(\lambda,m) &= 
\beta_0 + \beta_1 \log(\lambda) +\beta_2n + \beta_3 n^2 +\beta_4 n^3 + \beta_5 \lambda + \beta_6\lambda^2+ \beta_7\lambda^3 + \beta_8\log(n) + \beta_9\log(\lambda)n \\ &+ \beta_{10}\log(\lambda)n^2+ \beta_{11}\log(\lambda)n^3 + \beta_{12}n\lambda + \beta_{13} n^2\lambda + \beta_{14}n^3\lambda + \beta_{15}\lambda^4 + \beta_{15}\lambda \log(n) \\ & + \beta_{16}\lambda^5 + \beta_{17}\lambda^2\log(n) + \beta_{18}\lambda^3\log(n) + \text{error}\\
\end{align*}


The above model estimates the thresholds for the $\text{GEWMA}_t$ for homogeneous counts and obtained an in-control ATS of $100 \pm 7$ for $100 \le n \le 1000$, $0.01 \le \lambda \le 1$ and $\alpha = 0.10$.
The standard error of the model was 0.0007 and the correlation between the model fitted values and the corresponding actual simulated $h_D(\lambda)$ values was 0.99999.


\subsection*{Simulation Study Results}
Below, we provide tables for the simulation results described in Section \ref{sec:simulations}. 
\begin{sidewaystable}
  \centering 
\caption {Collaborative team ATS performance for $\text{GEWMA}_t$ with $n_{\Omega_t} = 6$}
\begin{tabular}{|c|r r r r r r r r| r r r r r r r r|} \hline
 & \multicolumn{16}{|c|}{Communication outbreaks in team of size 6 from a network of size 100}\\ \hline
 $\lambda$&\multicolumn{8}{|c|}{0.2}&\multicolumn{8}{|c|}{0.7}\\ \hline
 $k$&0.05&0.1&0.15&0.2&0.25&0.3&0.35&0.4&0.05&0.1&0.15&0.2&0.25&0.3&0.35&0.4\\
 \hline
$\delta$&\multicolumn{16}{|c|}{\bf ATS}\\ \hline
$0.5$&70.60&75.33&76.30&74.83&74.11&72.61&53.71&43.36&55.29&57.67&58.23&55.59&47.71&33.97&23.70&17.41\\
 $1.0$&50.25&56.20&57.65&48.34&45.86&34.02&20.27&16.08&27.69&33.63&37.48&34.24&24.05&14.30&9.19&6.59\\
$2.0$&28.15&33.89&34.26&28.32&19.64&12.62&8.05&6.16&15.57&18.32&19.46&16.79&11.60&6.60&4.29&3.09\\
 $3.0$&19.12&23.48&23.51&19.35&12.67&7.87&5.20&3.95&10.32&12.23&12.96&11.01&7.77&4.50&2.89&2.12\\
 $4.0$&14.70&15.10&17.46&14.37&9.91&5.78&3.74&2.94&8.23&9.57&10.18&8.22&5.69&3.52&2.24&1.80\\
 $5.0$&12.06&14.30&13.88&11.30&7.56&4.68&3.13&2.39&6.70&7.89&8.17&6.60&4.69&2.91&1.88&1.41\\
 $6.0$&10.31&12.15&12.14&9.35&6.29&3.96&2.60&2.11&5.72&6.68&7.03&5.68&3.89&2.46&1.69&1.07\\
$7.0$&8.97&10.55&10.33&7.77&5.43&3.41&2.28&1.89&5.22&5.84&5.96&4.83&3.37&2.13&1.46&1.00\\ 
$8.0$&8.05&9.41&9.30&7.25&4.81&3.01&2.01&1.76&4.40&5.26&5.40&4.34&3.02&1.94&1.21&1.00\\\hline
\end{tabular}
\end{sidewaystable}

\pagebreak

\begin{sidewaystable}
  \centering 
\caption {Collaborative team ATS performance for $\text{GEWMA}_t$ with $n_{\Omega_t} = 6$ pt. 2}
\begin{tabular}{|c|r r r r r| r r r r r| r r r r r |} \hline
 & \multicolumn{15}{|c|}{Communication outbreaks in team of size 6 from a network of size 100}\\ \hline
 $\lambda$&\multicolumn{5}{|c|}{0.2}&\multicolumn{5}{|c|}{0.4}&\multicolumn{5}{|c|}{0.7}\\ \hline
 $k$&0.4&0.45&0.5&{\bf 0.6}&0.7&0.4&0.45&0.5&{\bf 0.6}&0.7&0.4&0.45&0.5&{\bf 0.6}&0.7\\
 \hline
$\delta$&\multicolumn{15}{|c|}{\bf ATS}\\ \hline
$0.5$&43.36&42.98&41.90&{\bf 39.75}&50.45&43.36&24.01&21.39&{\bf 20.42}&21.61&17.41&14.30&12.98&{\bf 12.23}&13.64\\
 $1.0$&16.08&12.91&11.87&{\bf 11.62}&12.64&9.28&7.78&7.24&{\bf 6.93}&7.74&6.59&5.50&5.10&{\bf 5.02}&5.56\\
$2.0$&6.16&5.29&4.91&{\bf 4.70}&5.40&4.17&3.55&3.38&{\bf 3.28}&3.65&3.09&2.71&2.51&{\bf 2.49}&2.71\\
 $3.0$&3.95&3.37&3.18&{\bf 3.11}&3.52&2.74&2.42&2.27&{\bf 2.23}&2.47&2.12&1.92&1.84&{\bf 1.83}&1.96\\
 $4.0$&2.94&2.63&2.46&{\bf 2.43}&2.66&2.16&1.95&{\bf 1.81}&1.82&1.95&1.80&1.55&{\bf 1.39}&{\bf 1.39}&1.58\\
 $5.0$&2.39&2.13&2.06&{\bf 2.03}&2.18&1.87&1.72&1.57&{\bf 1.52}&1.68&1.41&1.10&{\bf 1.06}&{\bf 1.06}&1.19\\
 $6.0$&2.11&1.89&1.84&{\bf 1.75}&1.86&1.66&1.32&{\bf 1.19}&1.21&1.40&1.07&1.02&{\bf 1.00}&{\bf 1.00}&1.03\\
$7.0$&1.89&1.68&1.60&{\bf 1.56}&1.72&1.31&1.08&{\bf 1.05}&{\bf 1.05}&1.14&1.00&1.00&1.00&1.00&1.00\\ 
$8.0$&1.76&1.46&1.36&{\bf 1.35}&1.54&1.11&1.02&{\bf 1.00}&{\bf 1.00}&1.04&1.00&1.00&1.00&1.00&1.00\\\hline
\end{tabular}
\end{sidewaystable}

\pagebreak
\begin{sidewaystable}
  \centering 
\caption {Collaborative team ATS performance for $\text{GEWMA}_t$ with $n_{\Omega_t} = 7$}
\begin{tabular}{|c|r r r r r| r r r r r| r r r r r |} \hline
 & \multicolumn{15}{|c|}{Communication outbreaks in team of size 7 from a network of size 100}\\ \hline
 $\lambda$&\multicolumn{5}{|c|}{0.2}&\multicolumn{5}{|c|}{0.4}&\multicolumn{5}{|c|}{0.7}\\ \hline
$\lambda$&TEWMA&\multicolumn{4}{|c|}{GEWMA}&TEWMA&\multicolumn{4}{|c|}{GEWMA}&TEWMA&\multicolumn{4}{|c|}{GEWMA}\\ \hline
 $k$&&0.4&{\bf 0.5}&0.6&0.7&&0.4&0.5&{\bf 0.6}&0.7&&0.4&{\bf 0.5}&{\bf 0.6}&0.7\\
 \hline
$\delta$&\multicolumn{15}{|c|}{\bf ATS}\\ \hline
$0.25$&&&&&&&&&&&57.26&45.36&40.46&{\bf 38.98}&43.89\\
$0.5$&53.45&41.97&34.38&{\bf 32.10}&42.06&43.36&30.71&17.07&{\bf 16.95}&19.56&34.25&13.34&10.70&{\bf 10.69}&12.26\\
 $1.0$&31.84&12.16&{\bf 9.83}&9.96&11.58&22.00&9.73&{\bf 6.12}&6.16&6.90&16.08&5.44&{\bf 4.48}&4.52&5.02\\
$2.0$&14.93&4.986&{\bf 4.26}&4.26&4.80&9.65&4.16&2.98&{\bf 2.96}&3.30&6.68&2.65&{\bf 2.25}&2.30&2.54\\
 $3.0$&8.99&3.29&{\bf 2.86}&2.91&3.16&5.89&2.81&2.08&{\bf 2.06}&2.26&4.23&1.92&{\bf 1.72}&1.74&1.83\\
 $4.0$&6.25&2.57&{\bf 2.18}&2.22&2.47&4.19&2.11&1.72&{\bf 1.72}&1.84&3.08&1.53&{\bf 1.18}&1.22&1.33\\
 $5.0$&4.83&2.06&{\bf 1.68}&1.78&2.01&3.31&1.88&1.36&{\bf 1.34}&1.59&2.49&1.07&{\bf 1.01}&1.02&1.09\\
 $6.0$&3.95&1.83&{\bf 1.37}&1.39&1.79&2.75&1.63&{\bf 1.06}&1.08&1.24&2.12&1.00&1.00&1.00&1.01\\
$7.0$&3.36&1.68&{\bf 1.15}&1.20&1.59&2.39&1.34&{\bf 1.00}&1.01&1.06&1.89&1.00&1.00&1.00&1.00\\ 
$8.0$&2.94&1.45&{\bf 1.02}&1.05&1.10&2.11&1.10&1.00&1.00&1.00&1.39&1.00&1.00&1.00&1.00\\\hline
\end{tabular}
\end{sidewaystable}

\pagebreak
\begin{sidewaystable}
  \centering 
\caption {Collaborative team ATS performance for $\text{GEWMA}_t$ with $n_{\Omega_t} = 8$}
\begin{tabular}{|c|r r r r r| r r r r r| r r r r r |} \hline
 & \multicolumn{15}{|c|}{Communication outbreaks in team of size 8 from a network of size 100}\\ \hline
 $\lambda$&\multicolumn{5}{|c|}{0.2}&\multicolumn{5}{|c|}{0.4}&\multicolumn{5}{|c|}{0.7}\\ \hline
&TEWMA&\multicolumn{4}{|c|}{GEWMA}&TEWMA&\multicolumn{4}{|c|}{GEWMA}&TEWMA&\multicolumn{4}{|c|}{GEWMA}\\ \hline
 $k$&&0.4&{\bf 0.5}&0.6&0.7&&0.4&{\bf 0.5}&0.6&0.7&&0.4&{\bf 0.5}&0.6&0.7\\
 \hline
$\delta$&\multicolumn{15}{|c|}{\bf ATS}\\ \hline
$0.25$&&&&&&&&&&&48.55&{\bf 30.51}&31.99&33.17&38.11\\
$0.5$&44.73&30.51&27.60&{\bf 27.06}&32.37&34.98&16.79&{\bf 14.07}&14.50&17.12&25.38&9.89&{\bf 9.26}&9.54&10.91\\
 $1.0$&24.32&9.89&{\bf 8.60}&8.70&10.34&16.21&6.38&{\bf 5.60}&5.71&6.42&11.20&4.36&{\bf 4.05}&4.13&4.60\\
$2.0$&10.44&4.36&{\bf 3.87}&3.95&4.49&6.90&3.30&{\bf 2.71}&2.73&3.11&4.89&2.92&{\bf 2.09}&2.14&2.41\\
 $3.0$&6.36&2.92&2.62&{\bf 2.61}&2.97&4.33&2.13&{\bf 1.93}&1.94&2.15&3.19&2.20&{\bf 1.55}&1.62&1.77\\
 $4.0$&4.58&2.20&{\bf 2.04}&2.10&2.33&3.19&1.78&{\bf 1.58}&1.59&1.73&2.38&1.88&{\bf 1.06}&1.11&1.33\\
 $5.0$&3.62&1.88&{\bf 1.77}&1.79&1.91&2.55&1.37&{\bf 1.14}&1.18&1.41&2.01&1.69&{\bf 1.00}&1.00&1.04\\
 $6.0$&3.02&1.69&{\bf 1.49}&1.50&1.69&2.15&1.09&{\bf 1.03}&{\bf 1.03}&1.14&1.68&1.43&1.00&1.00&1.00\\
$7.0$&2.57&1.43&{\bf 1.20}&1.25&1.47&1.88&1.01&1.00&1.00&1.02&1.52&1.19&1.00&1.00&1.00\\ 
$8.0$&2.29&1.19&{\bf 1.06}&1.09&1.24&1.71&1.00&1.00&1.00&1.00&1.34&1.02&1.00&1.00&1.00\\\hline
\end{tabular}
\end{sidewaystable}

\clearpage

\begin{sidewaystable}
  \centering 
\caption {Collaborative team ATS performance for $\text{GEWMA}_t$ with $n_{\Omega_t} = 9$}
\begin{tabular}{|c|r r r r r| r r r r r| r r r r r |} \hline
 & \multicolumn{15}{|c|}{Communication outbreaks in team of size 9 from a network of size 100}\\ \hline
 $\lambda$&\multicolumn{5}{|c|}{0.2}&\multicolumn{5}{|c|}{0.4}&\multicolumn{5}{|c|}{0.7}\\ \hline
&TEWMA&\multicolumn{4}{|c|}{GEWMA}&TEWMA&\multicolumn{4}{|c|}{GEWMA}&TEWMA&\multicolumn{4}{|c|}{GEWMA}\\ \hline
 $k$&&0.4&{\bf 0.5}&0.6&0.7&&0.4&0.5&{\bf 0.6}&0.7&&0.4&0.5&{\bf 0.6}&0.7\\
 \hline
$\delta$&\multicolumn{15}{|c|}{\bf ATS}\\ \hline
$0.25$&&&&&&&&&&&38.20&28.93&25.89&26.56&33.81\\
$0.5$&38.23&24.63&{\bf 22.25}&23.50&30.47&27.38&12.45&{\bf 12.34}&13.17&15.19&20.06&9.07&{\bf 8.54}&8.62&9.97\\
 $1.0$&18.64&8.29&{\bf 7.58}&8.00&9.50&12.01&4.93&{\bf 4.86}&5.14&6.01&8.54&3.94&{\bf 3.71}&3.86&4.07\\
$2.0$&7.91&3.83&{\bf 3.53}&3.65&4.15&5.27&2.48&{\bf 2.47}&2.58&2.91&3.87&2.06&{\bf 1.90}&2.02&2.14\\
 $3.0$&4.89&2.56&{\bf 2.45}&2.46&2.82&3.39&1.82&{\bf 1.80}&1.87&2.01&2.52&1.54&{\bf 1.38}&1.48&1.77\\
 $4.0$&3.56&2.04&{\bf 1.89}&1.95&2.17&2.52&{\bf 1.40}&{\bf 1.40}&1.47&1.70&2.00&1.05&{\bf 1.01}&1.05&1.33\\
 $5.0$&2.84&1.80&{\bf 1.63}&1.66&1.81&2.05&{\bf 1.05}&{\bf 1.05}&1.09&1.33&1.64&1.00&1.00&1.00&1.04\\
 $6.0$&2.39&1.51&{\bf 1.30}&1.40&1.56&1.76&{\bf 1.00}&{\bf 1.00}&1.01&1.08&1.44&1.00&1.00&1.00&1.00\\
$7.0$&2.08&1.21&{\bf 1.07}&1.14&1.38&1.58&1.00&1.00&1.00&1.00&1.27&1.00&1.00&1.00&1.00\\ 
$8.0$&1.86&1.07&{\bf 1.02}&1.03&1.14&1.43&1.00&1.00&1.00&1.00&1.09&1.00&1.00&1.00&1.00\\\hline
\end{tabular}
\end{sidewaystable}

\clearpage
\begin{sidewaystable}
  \centering 
\caption {Collaborative team ATS performance for $\text{GEWMA}_t$ with $n_{\Omega_t} = 10$}
\begin{tabular}{|c|r r r r r| r r r r r| r r r r r |} \hline
 & \multicolumn{15}{|c|}{Communication outbreaks in team of size 10 from a network of size 100}\\ \hline
 $\lambda$&\multicolumn{5}{|c|}{0.2}&\multicolumn{5}{|c|}{0.4}&\multicolumn{5}{|c|}{0.7}\\ \hline
&TEWMA&\multicolumn{4}{|c|}{GEWMA}&TEWMA&\multicolumn{4}{|c|}{GEWMA}&TEWMA&\multicolumn{4}{|c|}{GEWMA}\\ \hline
 $k$&&0.4&{\bf 0.5}&0.6&0.7&&0.4&{\bf 0.5}&0.6&0.7&&0.4&{\bf 0.5}&0.6&0.7\\
 \hline
$\delta$&\multicolumn{15}{|c|}{\bf ATS}\\ \hline
$0.25$&&&&&&52.79&43.36&{\bf 37.06}&43.40&50.28&35.88&23.48&{\bf 22.57}&23.55&28.32\\
$0.50$&30.94&21.23&{\bf 19.82}&21.63&25.36&21.69&12.45&{\bf 10.84}&11.64&15.91&10.80&8.30&{\bf 7.27}&7.89&8.89\\
 $1.00$&14.38&7.25&{\bf 7.02}&7.33&8.60&9.42&4.93&{\bf 4.67}&4.80&6.63&4.62&3.64&{\bf 3.43}&3.66&4.07\\
$2.00$&6.12&3.39&{\bf 3.26}&3.47&3.87&4.12&2.47&{\bf 2.35}&2.45&2.81&3.03&1.94&{\bf 1.90}&1.96&2.14\\
 $3.00$&3.89&2.37&{\bf 2.20}&2.39&2.63&2.70&1.82&{\bf 1.76}&1.81&1.93&2.09&1.35&{\bf 1.23}&1.38&1.63\\
 $4.00$&2.88&1.90&{\bf 1.83}&1.86&1.99&2.09&1.40&{\bf 1.27}&1.38&1.65&1.26&{\bf 1.00}&{\bf 1.00}&1.02&1.14\\
 $5.00$&2.32&1.64&{\bf 1.51}&1.62&1.77&1.73&1.04&{\bf 1.02}&1.04&1.23&1.41&1.01&1.00&1.00&1.00\\
 $6.00$&1.99&1.30&{\bf 1.18}&1.28&1.49&1.51&1.00&1.00&1.00&1.04&1.22&1.00&1.00&1.00&1.00\\
$7.00$&1.75&1.06&{\bf 1.03}&1.09&1.25&1.35&1.00&1.00&1.00&1.00&1.08&1.00&1.00&1.00&1.00\\ 
$8.00$&1.57&1.04&{\bf 1.00}&1.02&1.09&1.22&1.00&1.00&1.00&1.00&1.00&1.00&1.00&1.00&1.00\\\hline
\end{tabular}
\end{sidewaystable}

\goodbreak

\begin{sidewaystable}
  \centering 
\caption {Collaborative team ATS performance for $\text{DEWMA}_{\nu,t}$ with $n_{\Omega_t} = 6$}
\begin{tabular}{|c|r r r| r r r | r r r |} \hline
 & \multicolumn{9}{|c|}{Communication outbreaks in team of size 6 from a network of size 100}\\ \hline
 $\lambda$&\multicolumn{3}{|c|}{0.2}&\multicolumn{3}{|c|}{0.4}&\multicolumn{3}{|c|}{0.7}\\ \hline
 $k$&\phantom{00}0.4\phantom{00}&\phantom{00}0.45\phantom{00}&\phantom{00}0.5\phantom{00}&\phantom{00}0.4\phantom{00}&\phantom{00}0.45\phantom{00}&\phantom{00}0.5\phantom{00}&\phantom{00}0.4\phantom{00}&\phantom{00}0.45\phantom{00}&\phantom{00}0.5\phantom{00}\\
 \hline
$\delta$&\multicolumn{9}{|c|}{\bf ATS}\\ \hline
$0.5$&&&&25.14&22.27&&14.94&13.50&14.57\\
 $1.0$&13.80&{\bf 12.90}&13.56&8.02&{\bf 7.78}&8.0&5.78&{\bf 5.25}&5.69\\
$2.0$&5.51&{\bf 5.34}&5.50&3.60&{\bf 3.57}&3.74&2.80&{\bf 2.66}&2.78\\
 $3.0$&360&{\bf 3.37}&3.55&2.49&{\bf 2.41}&2.51&1.99&{\bf 1.92}&1.97\\
 $4.0$&2.75&{\bf 2.55}&2.63&1.99&{\bf 1.91}&1.98&1.65&{\bf 1.60}&1.61\\
 $5.0$&2.23&{\bf 2.12}&2.20&1.71&{\bf 1.61}&1.69&1.15&{\bf 1.14}&1.20\\
 $6.0$&1.98&{\bf 1.91}&1.94&1.42&{\bf 1.38}&1.40&{\bf 1.01}&{\bf 1.01}&1.03\\
$7.0$&1.75&{\bf 1.71}&1.75&1.19&{\bf 1.18}&{\bf 1.18}&1.00&1.00&1.00\\ 
$8.0$&1.58&{\bf 1.51}&1.54&1.04&{\bf 1.03}&{\bf 1.03}&1.00&1.00&1.00\\\hline
\end{tabular}
\end{sidewaystable}

\pagebreak

\begin{sidewaystable}
  \centering 
\caption {Collaborative team ATS performance for $\text{DEWMA}_{\nu,t}$ with $n_{\Omega_t} = 7$}
\begin{tabular}{|c|r r r| r r r | r r r |} \hline
 & \multicolumn{9}{|c|}{Communication outbreaks in team of size 7 from a network of size 100}\\ \hline
 $\lambda$&\multicolumn{3}{|c|}{0.2}&\multicolumn{3}{|c|}{0.4}&\multicolumn{3}{|c|}{0.7}\\ \hline
 $k$&\phantom{00}0.4\phantom{00}&\phantom{00}0.45\phantom{00}&\phantom{00}0.5\phantom{00}&\phantom{00}0.4\phantom{00}&\phantom{00}0.45\phantom{00}&\phantom{00}0.5\phantom{00}&\phantom{00}0.4\phantom{00}&\phantom{00}0.45\phantom{00}&\phantom{00}0.5\phantom{00}\\
 \hline
$\delta$&\multicolumn{9}{|c|}{\bf ATS}\\ \hline
$0.5$&&&38.72&18.44&{\bf 16.89}&19.57&11.97&{\bf 11.36}&12.56\\
 $1.0$&10.56&{\bf 10.52}&11.18&6.72&{\bf 6.66}&7.42&{\bf 4.78}&4.82&5.57\\
$2.0$&4.67&{\bf 4.61}&4.90&3.20&{\bf 3.19}&3.44&{\bf 2.44}&2.46&2.76\\
 $3.0$&3.12&{\bf 3.09}&3.21&2.26&{\bf 2.18}&2.36&{\bf 1.84}&1.86&1.96\\
 $4.0$&2.43&{\bf 2.35}&2.47&{\bf 1.85}&{\bf 1.85}&1.91&1.33&{\bf 1.29}&1.60\\
 $5.0$&1.96&{\bf 1.93}&1.99&{\bf 1.45}&{\bf 1.45}&1.60&{\bf 1.04}&{\bf 1.04}&1.20\\
 $6.0$&1.83&{\bf 1.73}&1.81&1.17&{\bf 1.14}&1.33&1.01&{\bf 1.00}&1.04\\
$7.0$&1.59&{\bf 1.54}&1.62&1.04&{\bf 1.03}&1.13&1.00&1.00&1.00\\ 
$8.0$&1.36&{\bf 1.32}&1.40&{\bf 1.00}&{\bf 1.00}&1.04&1.00&1.00&1.00\\\hline
\end{tabular}
\end{sidewaystable}

\begin{sidewaystable}
  \centering 
\caption {Collaborative team ATS performance for $\text{DEWMA}_{\nu,t}$ with $n_{\Omega_t} = 8$}
\begin{tabular}{|c|r r r| r r r | r r r |} \hline
 & \multicolumn{9}{|c|}{Communication outbreaks in team of size 7 from a network of size 100}\\ \hline
 $\lambda$&\multicolumn{3}{|c|}{0.2}&\multicolumn{3}{|c|}{0.4}&\multicolumn{3}{|c|}{0.7}\\ \hline
 $k$&\phantom{00}0.4\phantom{00}&\phantom{00}0.45\phantom{00}&\phantom{00}0.5\phantom{00}&\phantom{00}0.4\phantom{00}&\phantom{00}0.45\phantom{00}&\phantom{00}0.5\phantom{00}&\phantom{00}0.4\phantom{00}&\phantom{00}0.45\phantom{00}&\phantom{00}0.5\phantom{00}\\
 \hline
$\delta$&\multicolumn{9}{|c|}{\bf ATS}\\ \hline
$0.25$&&&&&&&32.18&35.76&37.24\\
$0.5$&32.78&32.88&33.92&15.66&{\bf 15.29}&16.32&{\bf 10.07}&10.16&10.34\\
 $1.0$&{\bf 9.54}&9.64&10.26&{\bf 6.09}&6.20&6.32&4.36&{\bf 4.24}&4.62\\
$2.0$&{\bf 4.14}&4.17&4.32&{\bf 2.84}&2.90&3.02&2.42&{\bf 2.31}&2.35\\
 $3.0$&2.78&{\bf 2.74}&2.94&{\bf 2.01}&2.03&2.14&1.70&{\bf 1.68}&1.76\\
 $4.0$&2.12&{\bf 2.10}&2.35&1.66&{\bf 1.65}&1.78&1.21&{\bf 1.19}&1.28\\
 $5.0$&{\bf 1.86}&{\bf 1.86}&1.90&1.32&{\bf 1.28}&1.40&{\bf 1.01}&{\bf 1.01}&1.02\\
 $6.0$&1.68&{\bf 1.63}&1.71&1.05&{\bf 1.04}&1.09&1.00&1.00&1.00\\
$7.0$&1.37&{\bf 1.36}&1.50&{\bf 1.00}&{\bf 1.00}&1.01&1.00&1.00&1.00\\\hline
\end{tabular}
\end{sidewaystable}
\clearpage

\begin{sidewaystable}
  \centering 
\caption {Collaborative team ATS performance for $\text{DEWMA}_{\nu,t}$ with $n_{\Omega_t} = 9$}
\begin{tabular}{|c|r r r| r r r | r r r |} \hline
 & \multicolumn{9}{|c|}{Communication outbreaks in team of size 9 from a network of size 100}\\ \hline
 $\lambda$&\multicolumn{3}{|c|}{0.2}&\multicolumn{3}{|c|}{0.4}&\multicolumn{3}{|c|}{0.7}\\ \hline
 $k$&\phantom{00}0.4\phantom{00}&\phantom{00}0.45\phantom{00}&\phantom{00}0.5\phantom{00}&\phantom{00}0.4\phantom{00}&\phantom{00}0.45\phantom{00}&\phantom{00}0.5\phantom{00}&\phantom{00}0.4\phantom{00}&\phantom{00}0.45\phantom{00}&\phantom{00}0.5\phantom{00}\\
 \hline
$\delta$&\multicolumn{9}{|c|}{\bf ATS}\\ \hline
$0.25$&&&&54.90&{\bf 48.93}&52.01&28.93&{\bf 26.74}&29.78\\
$0.5$&22.27&21.86&25.7&14.26&{\bf 13.10}&14.06&{\bf 8.87}&8.94&9.33\\
 $1.0$&8.47&{\bf 8.14}&9.05&5.51&{\bf 5.38}&5.76&{\bf 3.91}&4.04&4.16\\
$2.0$&3.81&{\bf 3.80}&3.96&2.73&{\bf 2.60}&2.84&{\bf 2.14}&2.15&2.19\\
 $3.0$&{\bf 2.59}&{\bf 2.59}&2.84&1.94&{\bf 1.91}&2.01&{\bf 1.52}&1.58&1.70\\
 $4.0$&{\bf 2.04}&{\bf 2.04}&2.13&{\bf 1.48}&{\bf 1.48}&1.68&{\bf 1.04}& 1.10&1.13\\
 $5.0$&1.80&{\bf 1.79}&1.86&{\bf 1.19}&{\bf 1.19}&1.26&1.00&1.00&1.00\\
 $6.0$&1.49&{\bf 1.48}&1.62&{\bf 1.03}&{\bf 1.03}&1.02&1.00&1.00&1.00\\
$7.0$&1.25&{\bf 1.19}&1.32&1.00&1.00&1.00&1.00&1.00&1.00\\
$8.0$&1.09&{\bf 1.06}&1.10&1.00&1.100&1.00&1.00&1.00&1.00\\\hline
\end{tabular}
\end{sidewaystable}


\begin{sidewaystable}
\centering 
\caption {Dominant leader team outbreaks involving teams of size 6 to 9}
\begin{tabular}{|c|r r r r | r r r r | r r r r  | r r r r  |} \hline
 & \multicolumn{4}{|c|}{DEWMA}&\multicolumn{4}{|c|}{GEWMA}&\multicolumn{4}{|c|}{DEWMA}&\multicolumn{4}{|c|}{GEWMA}\\ \hline
 $\lambda$&\multicolumn{8}{|c|}{0.2}&\multicolumn{8}{|c|}{0.4}\\ \hline
$n_{\Omega}$&6&7&8&9&6&7&8&9&6&7&8&9&6&7&8&9\\ \hline
$k$&\multicolumn{4}{|c|}{\bf 0.45}&\multicolumn{4}{|c|}{\bf 0.6}&\multicolumn{4}{|c|}{\bf 0.45}&\multicolumn{4}{|c|}{\bf 0.6}\\ \hline
$\delta$&\multicolumn{16}{|c|}{\bf ATS}\\ \hline
$0.25$&&&&&&&&&&&62.66&54.82&&&98.60&83.27\\
$0.50$&49.48&39.94&34.19&33.78&72.00&63.46&44.72&39.01&27.42&22.54&20.24&15.94&32.74&24.59&21.99&17.97\\
 $1.00$&15.99&13.19&11.29&10.51&16.91&14.32&13.24&11.44&9.99&8.49&7.27&6.59&9.03&8.42&7.42&6.73\\
$2.00$&6.26&5.65&5.09&4.62&6.27&5.68&5.43&4.65&4.29&3.86&3.46&3.16&4.18&3.73&3.67&3.16\\
 $3.00$&4.25&3.77&3.31&2.39&4.17&3.68&3.48&3.00&2.90&2.56&2.42&2.19&2.83&2.59&2.42&2.21\\
 $4.00$&3.20&2.83&2.62&1.73&3.18&2.73&2.66&2.38&2.28&2.09&1.90&1.70&2.17&2.06&1.92&1.79\\
 $5.00$&2.66&2.28&2.11&1.32&2.55&2.39&2.22&1.95&1.94&1.70&1.56&1.36&1.84&1.71&1.59&1.42\\
 $6.00$&2.16&2.00&1.64&1.06&2.19&2.06&1.86&1.74&1.65&1.48&1.32&1.16&1.61&1.48&1.30&1.18\\
$7.00$&1.89&1.78&1.43&1.00&2.01&1.72&1.67&1.52&1.43&1.29&1.10&1.04&1.34&1.29&1.13&1.07\\ 
$8.00$&1.64&1.36&1.24&1.00&1.70&1.61&1.48&1.35&1.22&1.04&1.01&1.00&1.17&1.06&1.00&1.00\\\hline
\end{tabular}
\end{sidewaystable}

\begin{sidewaystable}
\centering  
\caption {Dominant leader team outbreaks involving teams of size 6 to 9}
\begin{tabular}{|c|r r r r | r r r r | } \hline
 & \multicolumn{4}{|c|}{DEWMA}&\multicolumn{4}{|c|}{GEWMA}\\ \hline
 $\lambda$&\multicolumn{8}{|c|}{0.7}\\ \hline
$n_{\Omega}$&6&7&8&9&6&7&8&9\\ \hline
$k$&\multicolumn{4}{|c|}{\bf 0.45}&\multicolumn{4}{|c|}{\bf 0.6} \\\hline
$\delta$&\multicolumn{8}{|c|}{\bf ATS}\\ \hline
$0.25$&53.61&46.84&40.94&33.88&98.12&80.02&67.48&44.12\\
$0.50$&16.95&13.82&12.72&10.51&18.12&15.48&14.72&10.94\\
 $1.00$&6.77&6.01&5.25&4.62&6.38&5.88&5.47&4.81\\
$2.00$&3.28&2.90&2.66&2.38&3.04&2.84&2.74&2.46\\
 $3.00$&2.28&2.03&1.93&1.73&2.17&1.98&2.00&1.83\\
 $4.00$&1.82&1.70&1.51&1.32&1.76&1.66&1.56&1.39\\
 $5.00$&1.53&1.21&1.16&1.06&1.40&1.31&1.19&1.09\\
 $6.00$&1.23&1.09&1.02&1.00&1.13&1.01&1.01&1.00\\
$7.00$&1.02&1.00&1.00&1.00&1.02&1.00&1.00&1.00\\ 
\hline
\end{tabular}
\end{sidewaystable}

\begin{sidewaystable}
\centering
\caption {The ADEWMA plans for heterogeneous networks with no outbreak}
\begin{tabular}{|c|c|c|c|c|} \hline
$m_{L}$ & $m_{H}$ & Threshold Adjustment & ATS & $a$\\ \hline
100&135&1.005&102.9&-0.0030\\
115&135&1.0037&103.1&-0.0030\\
110&150&0.982&104.0&-0.0060\\
130&150&0.984&102.2&-0.0060\\
100&175&0.984&100.9&-0.0050\\
115&175&0.984&102.2&-0.0050\\
135&175&0.982&103.3&-0.0050\\
155&175&0.982&101.6&-0.0050\\
135&250&0.985&99.4&-0.0035\\
200&250&0.983&102.4&-0.0035\\
150&275&0.985&100.9&-0.0030\\
215&275&0.985&103.6&-0.0030\\
305&315&0.981&101.6&-0.0027\\
250&350&0.986&99.3&-0.0025\\
\hline
\end{tabular}
\end{sidewaystable}
\clearpage

\bibliographystyle{Chicago}

\end{document}